 \documentclass[referee,a4paper,12pt,structabstract,english]{swsc2}
\usepackage[T1]{fontenc}
\usepackage{amstext}
\usepackage{amssymb,amsmath}
\usepackage{graphicx,txfonts,subfigure,lineno,url,multirow,array}
\usepackage[backref]{hyperref}
\usepackage[authoryear,round]{natbib}
\PassOptionsToPackage{normalem}{ulem}
\usepackage{ulem}

\hypersetup{colorlinks=true,citecolor=cyan,urlcolor=cyan,linkcolor=blue}

\title{Image patch analysis of sunspots and active regions. II. Clustering via matrix factorization}
\titlerunning{Image patch analysis of sunspots and active regions. II. Clustering}
\authorrunning{Moon et al}
\author{Kevin R. Moon\inst{1,*} \and V\'{e}ronique Delouille\inst{2} \and Jimmy J. Li\inst{1} \and Ruben De Visscher\inst{2} \and Fraser Watson\inst{3} \and Alfred O. Hero III\inst{1}}
\institute{Electrical Engineering and Computer Science Department, University of Michigan 
\\ *Corresponding author: \texttt{krmoon@umich.edu}  \and  SIDC, Royal Observatory of Belgium \and National Solar Observatory, Boulder, CO}
\usepackage{multirow}

\makeatother

\usepackage{babel}

\begin{document}
\abstract{Separating active regions that are quiet from potentially eruptive ones is a key issue in Space Weather applications. Traditional classification schemes such as Mount Wilson and McIntosh have been effective in relating an active region large scale magnetic configuration to its ability to produce eruptive events. However, their qualitative nature prevents systematic studies of an active region's evolution for example.}
{We introduce a new clustering of active regions that is based on the local geometry observed in Line of Sight magnetogram and continuum images.}
{We use a reduced-dimension representation of an active region that is obtained by factoring the corresponding data matrix comprised of local image patches. Two factorizations can be compared via the definition of appropriate metrics on the resulting factors. The distances obtained from these metrics are then used to cluster the active regions.}
{We find that these metrics result in natural clusterings of active regions. The clusterings are related to large scale descriptors of an active region such as its size, its local magnetic field distribution, and its complexity as measured by the Mount Wilson classification scheme. We also find that including data focused on the neutral line of an active region can result in an increased correspondence between our clustering results and other active region descriptors such as the Mount Wilson classifications and the $R$ value.}
{Matrix factorization of image patches is a promising new way of characterizing active regions. We provide some recommendations for which metrics, matrix factorization techniques, and regions of interest to use to study active regions.}
\maketitle

\section{Introduction}

\label{sec:intro}

\subsection{Context}
Identifying properties of active regions (AR) that
are necessary and sufficient for the production of energetic events
such as solar flares is one of the key issues in space weather. The
Mount Wilson classification (see Table~\ref{tab:mwilson_classification}
for a brief description of its four main classes) has been effective
in relating a sunspot's large scale magnetic configuration with its
ability to produce flares. \cite{1960AN....285..271K} pointed out
the first clear connection between flare productivity and magnetic
structure, and introduced a new magnetic classification, $\delta$,
to supplement Hale's $\alpha$, $\beta$ and $\gamma$ classes \citep{1919ApJ....49..153H}.
Several studies showed that a large proportion of all major flare
events begin with a $\delta$ configuration \citep{1966ApJ...145..215W,1985SoPh...96..293M,2000ApJ...540..583S}.

\begin{table}

\caption{Mount Wilson classification rules, number of each AR, and total number of joint patches or pixels per Mt. Wilson
class used in this paper when using the STARA masks. \label{tab:mwilson_classification}}
\centering

\begin{tabular}{|l|l|c|c|}
\hline 
Class & Classification Rule & Number of AR & Number of Patches \tabularnewline
\hline 
$\alpha$ & A single dominant spot & 50 & 13,358 \tabularnewline
\hline 
$\beta$ & A pair of dominant spots of opposite polarity & 192 & 75,463 \tabularnewline
\hline 
$\beta\gamma$ & A $\beta$ sunspot where a single north-south polarity & 130 & 95,631 \tabularnewline
 & inversion line cannot divide the two polarities & & \tabularnewline
\hline 
$\beta\gamma\delta$ & A $\beta\gamma$ sunspot where umbrae of opposite polarity  & 52 & 66,195 \tabularnewline
 & are together in a single penumbra & & \tabularnewline
\hline
\end{tabular}

\end{table}

The categorical nature of the Mount Wilson classification, however, prevents the differentiation between two sunspots with the same classification
and makes the study of an AR's evolution cumbersome. Moreover, the Mount Wilson classification is generally carried out manually which results in human bias. Several papers \citep{2013Stenning,2008SoPh..248..277C,2009SpWea...7.6001C} have used supervised techniques to reproduce the Mount Wilson and other schemes which has resulted in a reduction in human bias. 

To go beyond categorical classification in the flare prediction problem, 
the last decade has seen many efforts in describing the photospheric magnetic configuration
in more details. Typically, a set of scalar properties is derived
from line of sight (LOS) or vector magnetogram and analyzed in a supervised
classification context to derive which combination of properties is predictive of
increased flaring activity \citep{2004AAS...204.3905L,2006SoPh..237...25G,2007ApJ...655L.117S,2007SpWea...509002B,2007ApJ...661L.109G,2008ApJ...689.1433F,2009SoPh..254..101S,2010SoPh..263..175H,2010ApJ...710..869Y,2011SoPh..tmp..404A,2012SoPh..281..639L,2015ApJ...798..135B}.
Examples of scalar properties include: sunspot area, total unsigned
magnetic flux, flux imbalance, neutral line length, maximum gradients
along the neutral line, or other proxies for magnetic connectivity
within ARs. These scalar properties are features that can be used as input in flare prediction. However, there is no guarantee that these selected features exploit the information present in the data in an optimal way for the flare prediction problem.

\subsection{Contribution}

 We introduce a new data-driven method to cluster ARs using 
 information contained in magnetogram and continuum. Instead of focusing on the best set of properties that summarizes the information contained in those images, we study the natural geometry present in the data via a reduced-dimension representation of such images. The reduced-dimension is implemented via  matrix factorization of an image patch representation as explained in Section~\ref{S:matrix_fact}.
 We show how this geometry can be used for classifying  ARs in an unsupervised way, that is, without including AR labels as input to the analysis.

 We consider the same dataset as in~\cite{2015arXiv150304127M}. It is obtained from the \emph{Michelson Doppler Imager} (MDI) instrument~\citep{1995MDI}
 on board the SOHO Spacecraft. SOHO-MDI provides two to four times
 per day a white-light continuum image observed in the vicinity of
 the Ni~{\sc i} 676.7L~nm photospheric absorption line. MDI LOS magnetograms
 are recorded with a higher nominal cadence of 96 minutes. We selected
 424 sunspot images within the time range of 1996-2010. They span the
 various Mount Wilson classifications (see Table~\ref{tab:mwilson_classification}),
 are located within $30^\circ$ of central meridian, and have corresponding
 observations in both MDI continuum and MDI LOS magnetogram. We use
 level 1.8 data for both modalities.
 
\begin{figure}
\centering

\includegraphics[width=1\textwidth]{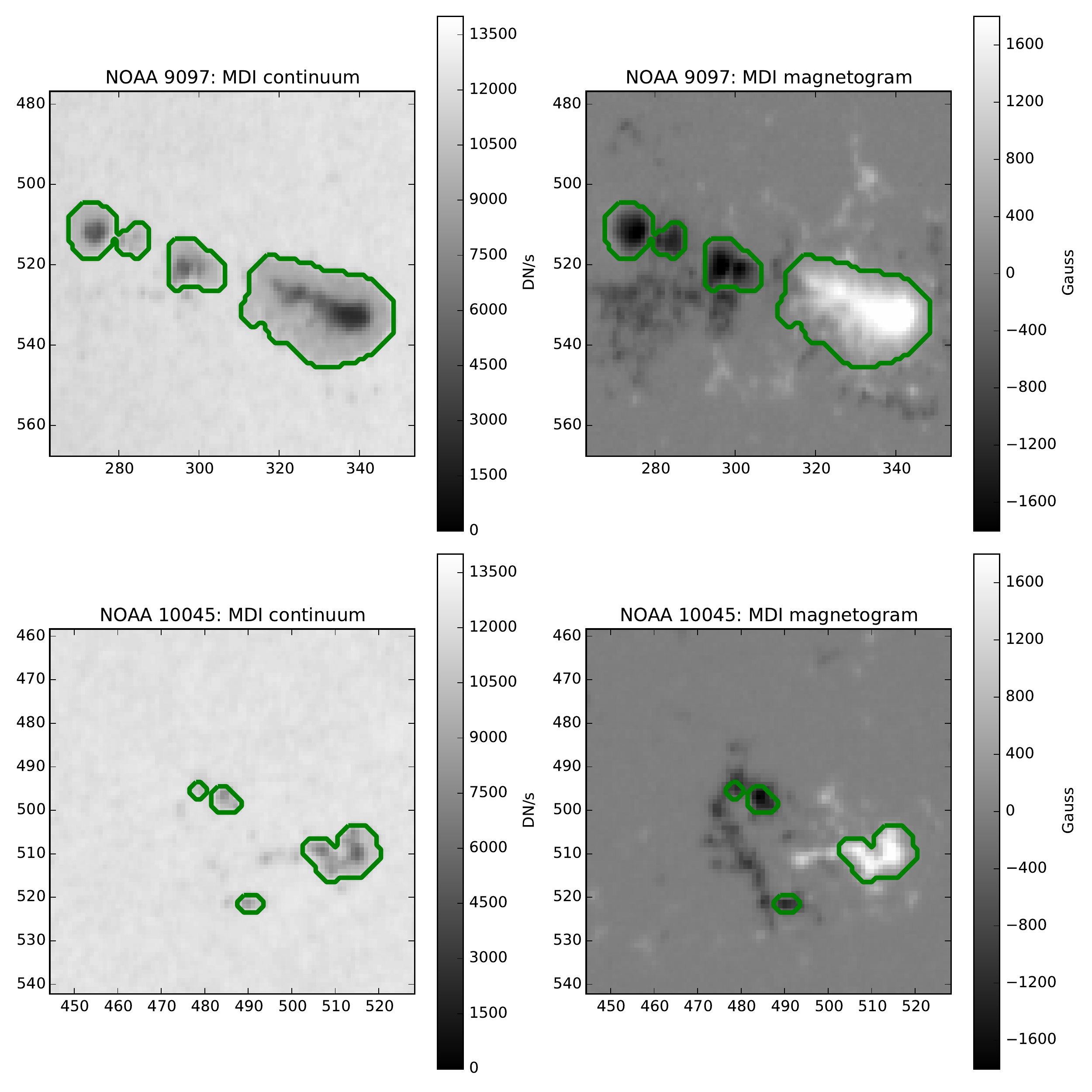}

\caption{MDI continuum and magnetogram from  NOAA 9097 on July 23, 2000 (top) and from NOAA 10045   on July 25, 2002 (bottom) overlaid with the
corresponding STARA masks in green. \label{fig:AR_examples}}

\end{figure}

 Our method can be adapted to any definition of the support of an AR, or Region of Interest (ROI), and such ROI must be given a priori. We consider three types of ROIs:
 \begin{enumerate}
\item  Umbrae and penumbrae masks obtained with the Sunspot Tracking and Recognition
 Algorithm (STARA) \citep{2011masks} from continuum
 images. These sunspot masks encompass the regions of highest variation observed in both continuum and magnetogram images, and hence are used primarily to illustrate our method. 
 Figure~\ref{fig:AR_examples} provides some examples of AR
 images overlaid with their respective STARA masks.
 \item The neutral line region, defined as the set of pixels situated no more than 10 pixels (20 arcsec) away from the neutral line, and located within the Solar Monitor
 Active Region Tracker (SMART) masks \citep{higgins2011solar}, which defines magnetic AR boundaries.  
 \item The set of pixels that are used as support for the computation of the $R$-value~defined in~\cite{2007ApJ...655L.117S}. The $R$-value measures a weighted absolute magnetic flux, where the weights are positive only around the neutral line. 
 \end{enumerate}
 
 Our patch-based matrix factorization method investigates the fine
  scale structures encoded by localized gradients of various directions
 and amplitudes, or locally smooth areas for example. In contrast,
 the Mount Wilson classification encodes the relative locations and
 sizes of concentrations of opposite polarity magnetic flux on a large
 scale. Although both classification schemes rely on completely different
 methods, using the first  ROI defined above, we find some similarities (see Section~\ref{sec:cluster_results}). Moreover the Mount Wilson classification
 can guide us in the interpretation of the results
 and clusters obtained. 
 

 The shape of the neutral line separating the
 two main polarities in an AR is a key element in the Mount Wilson
 classification scheme, and the magnetic field gradients observed along
 the neutral line are important information in the quest for solar
 activity prediction~\citep{2007ApJ...655L.117S,2008ApJ...689.1433F}. We  therefore
 analyze the effect of including the neutral line region in Section~\ref{sec:cluster_results}.
 
  Results based on the third ROI are compared directly to the $R$-value. The various comparisons enable us to evaluate the potential of our method for flare prediction.

\subsection{Reduced dimension via matrix factorization}
\label{S:matrix_fact}
Our data-driven method is based on a  reduced-dimension representation of an AR ROI via matrix factorization of image patches. Matrix factorization is a widely used tool to reveal patterns in high dimensional datasets. Applications  outside of solar physics are numerous and range e.g. from multimedia activity correlation, neuroscience, gene expression~\citep{bazot_gene}, to hyperspectral imaging~\citep{mittleman_hyper}. 

The idea is to express a $k-$multivariate observation $\mathbf{z}_1$ as a linear combination of a reduced number of $r < k$ components $\mathbf{a}_{j}$, each weighted by some (possibly random) coefficients $h_{j,1}$:
\begin{equation}
\label{E:z1}
\mathbf{z}_1 = \sum_{j=1}^r \mathbf{a}_{j} h_{j,1}~ + \mathbf{n}_1,
 \end{equation}
where $\mathbf{n}_1$ represents residual noise.
With $\mathbf{Z} = [\mathbf{z}_1,\ldots,\mathbf{z}_n]$, the equivalent matrix factorization representation is written as
\begin{equation}
\label{E:fact}
\mathbf{Z}= \mathbf{A}\mathbf{H} + \mathbf{N}
\end{equation}
where $\mathbf{Z}$  is a $k \times n$ data matrix containing $n$ observations of $k$ different variables, $\mathbf{A}$ is the $k\times r$ matrix containing the \lq dictionary elements' (called 'factor loadings' in some applications) and $\mathbf{H}$ is the $r\times n$
matrix of coefficients (or \lq factor scores'). The $k \times n$ matrix $\mathbf{N}$ contains residuals from the matrix  factorization model fitting.
Finding  $\mathbf{A}$ and  $\mathbf{H}$ from the knowledge of  $\mathbf{Z}$  alone  is a severely ill-posed problem, hence prior knowledge is needed
to constrain the solution to be unique. 

Principal Component Analysis (PCA)~\citep{Jolliffe2002} is probably the most widely used dimensionality reduction technique. It seeks principal directions that capture the highest variance in the data under the constraints that these directions are mutually orthogonal, thereby defining a subspace of the initial space that exhibit information rather than noise.
The PCA solution can be written as a matrix factorization thanks to the Singular Value Decomposition (SVD) \citep{moon2000methods}, and so we use SVD in the clustering method presented here. 

The Nonnegative Matrix Factorization (NMF) \citep{lee2001nmf} is also considered in this paper. Instead of imposing orthogonality, it constrains elements of matrices $\mathbf{A}$ and $\mathbf{H}$ to be nonnegative. We further impose that each column of $\mathbf{H}$  has elements that sum up to one, thereby effectively using a formulation identical to one used in hyperspectral unmixing \citep{Bioucas_IEEE_JSTARS_2012}. 

Unmixing techniques exploit the high redundancy observed in
similar bandpasses. They aim at separating the various  contributions and at estimating a smaller set of less dependent source images. Matrix factorization, known as  \lq blind source separation' in this context, has many applications, ranging from biomedical  imaging,  chemometrics, to  remote  sensing~\citep{comon_ica}, and recently to the extraction of salient morphological features from multi-wavelength extreme ultraviolet solar images~\citep{2013SoPh..283...31D}.
 
 In this paper, we wish to factorize  a  
 $k \times n$ data matrix $\mathbf{Z}$ containing $n$ observations
of $k$ different variables
 as in Equation~(\ref{E:fact}) where the dictionary matrix  $\mathbf{A}$ spans a subspace of the initial space, with $r<k$.
 We consider the cases where $\mathbf{Z}$ is formed from a single image as well as from multiple images. 

\begin{figure}
\centering

\includegraphics[width=0.4\textwidth]{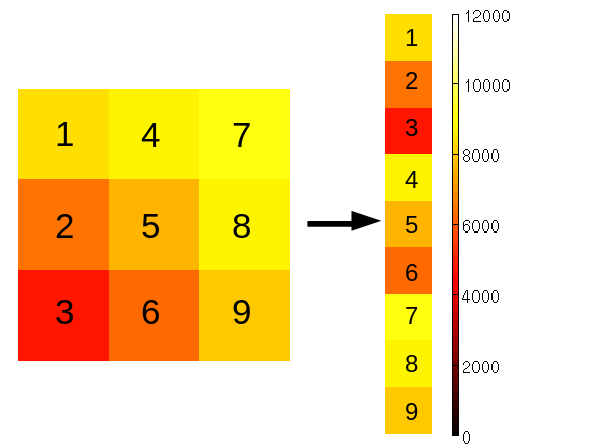}\includegraphics[width=0.5\textwidth]{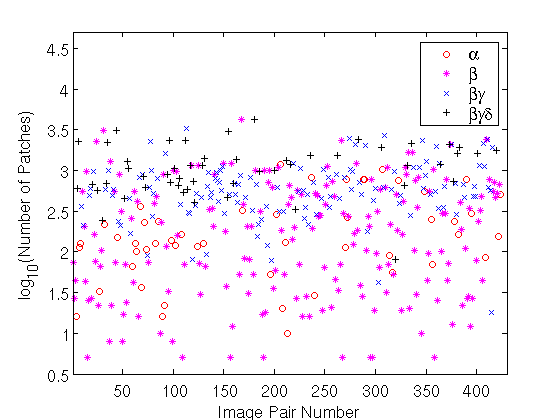}

\caption{(Left) An example of a $3\times 3$ pixel neighborhood or \emph{patch} extracted from the edge of a sunspot in a continuum image and
its column representation. (Right) The number of patches extracted from each pair of AR images when using the STARA masks. \label{fig:patch}}
\end{figure}

When a single image is used, the data
matrix $\mathbf{Z}$ is built from a $n$ pixel
image by taking overlapping $m \times m$-pixel neighborhoods called
\emph{patches}. Figure~\ref{fig:patch} (left)  presents such a patch and
its column representation. The $k$ rows of the $i-$th column of
$\mathbf{Z}$ are thus given by the $m^2$ pixel values in the neighborhood
of pixel $i$. The right plot in Figure~\ref{fig:patch} provides the number of patches in each pair of AR images when using the STARA masks. When multiple images are used, such as when analyzing collectively all images from a given Mount Wilson class, the patches are combined into a single data matrix. Table~\ref{tab:mwilson_classification} gives the total number of patches from each Mount Wilson class when using the STARA masks.  

\begin{figure}
\centering

\includegraphics[width=0.85\textwidth]{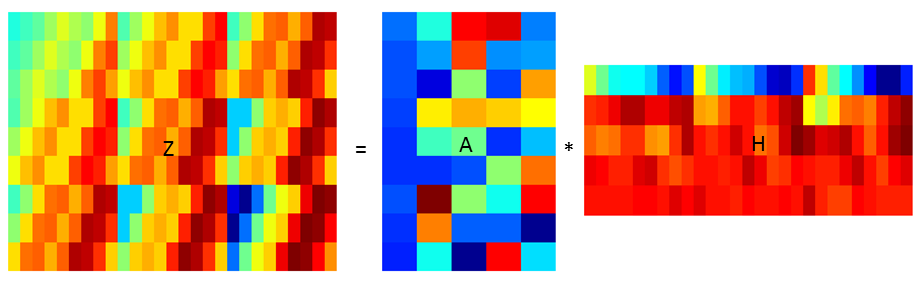}

\caption{An example of linear dimensionality reduction where the data matrix
of AR image patches $\mathbf{Z}$ is factored as a product of a dictionary
$\mathbf{A}$ of representative elements and the corresponding coefficients
in the matrix $\mathbf{H}$. The $\mathbf{A}$ matrix consists of
the basic building blocks for the data matrix $\mathbf{Z}$ and $\mathbf{H}$
contains the corresponding coefficients.\label{fig:matrix_factors}}
\end{figure}

A factorization of a data matrix containing
image patches is illustrated in Figure~\ref{fig:matrix_factors}. 
In this figure, let $\mathbf{z}_{1}$ be the first column of $\mathbf{Z}$, containing the  intensity values for the first patch. These intensity values are decomposed as a sum of $r$ elements as in Equation~(\ref{E:z1})
where $\mathbf{a}_{j}$ is the $j-$th column of $\mathbf{A}$ and $h_{j,1}$ is the $(j,1)-$th element of $\mathbf{H}$. In this representation, the vectors $\mathbf{a}_{j}, j=1,\ldots,r$ are the elementary building blocks common to all patches, whereas the $h_{j,1}$ are the coefficients specific to the first patch. 

To compare ARs and cluster them based on this reduced dimension representation, some form of distance is required. To measure the distance between two ARs, we apply some metrics to the corresponding matrices $\mathbf{A}$ or $\mathbf{H}$ obtained from the factorizations of the two ARs. These distances are further introduced into a clustering algorithm that groups ARs based on the similarity of their patch geometry.

This paper builds upon results obtained in \cite{2015arXiv150304127M}, which can be summarized as follows:
\begin{enumerate}
\item Continuum and magnetogram modalities are correlated, and there may
be some advantage in considering both of them in an analysis.
\item A patch size of $m=3$ includes a significant portion of spatial correlations
present in continuum and magnetogram images.
\item Linear methods for dimensionality reduction (e.g. matrix factorization) are appropriate to analyze
ARs observed with continuum images and magnetogram.
\end{enumerate}

With an AR area equal to $n$ pixels and an $m\times m$ patch, the
corresponding continuum data matrix $\mathbf{X}$ and magnetogram
data matrix $\mathbf{Y}$ each have size $m^{2}\times n$. The full
data matrix considered is $\mathbf{Z}=\left(\begin{array}{c}
\mathbf{X}\\
\mathbf{Y}\end{array}\right)$ with size $2m^{2}\times n=18\times n$ in our case. The images extracted from both modalities are normalized prior to analysis. An intrinsic
dimension analysis in \cite{2015arXiv150304127M} showed further that a sunspot can be
represented accurately with a dictionary containing six elements. 

\subsection{Outline}

Section~\ref{sec:dictionary} describes two matrix factorization methods: the singular value decomposition (SVD) and nonnegative matrix factorization (NMF). While more sophisticated methods exist that may lead to improved performance, we focus on SVD and NMF to demonstrate the utility of an analysis of a reduced dimension representation of image patches for this problem. Future work will include further refinement in the choice of matrix factorization techniques.
To compare the results from this factorization we need a metric,
and so we use the Hellinger
distance for this purpose. To obtain some insight on how these factorizations
separate the data, we make some general comparisons in Section~\ref{sec:dict_compare}.
In particular, with the defined metric, we compute the pairwise distances
between Mount Wilson classes to identify which classes are most similar
or dissimilar according to the matrix factorization results. 

Section~\ref{sec:cluster} describes the clustering procedures that
take the metrics' output as input. The method called \lq Evidence Accumulating
Clustering with Dual rooted Prim tree Cuts' (EAC-DC) was introduced
by \cite{galluccio2013clustering} and is used to cluster the ARs.
By combining the two matrix factorization methods,
a total of two procedures are used to analyze the data. Besides analyzing
the whole sunspot data, we also look at information contained in patches
situated along the neutral lines. The results of the clustering analyses
are provided in Section~\ref{sec:cluster_results}.

This paper improves and extends the work in \cite{moon2014icip},
where fixed size square pixel regions centered on the sunspot group
were used as the ROI for the matrix factorization prior to clustering.
Moreover, here we are using more appropriate metrics to compare the
factorization results.

\section{Matrix Factorization}

\label{sec:dictionary}The intrinsic dimension analysis in \cite{2015arXiv150304127M}
showed that linear methods (e.g. matrix factorization)  are sufficient to represent the data, and
hence we focus on those. Matrix factorization methods aim at
finding a set of basis vectors or dictionary elements such that each
data point (in our case, pair of pixel patches) can be accurately
expressed as a linear combination of the dictionary elements. Mathematically,
if we use $m\times m$ patches then this can be expressed as $\mathbf{Z}\approx\mathbf{AH}$,
where $\mathbf{Z}$ is the $2m^{2}\times n$ data matrix with $n$
data points being considered, $\mathbf{A}$ is the $2m^{2}\times r$
dictionary with the columns corresponding to the dictionary elements,
and \textbf{$\mathbf{H}$ }is the $r\times n$ matrix of coefficients.
The goal is to find matrices $\mathbf{A}$ and $\mathbf{H}$ whose
product nearly approximates $\mathbf{Z}$. The degree of approximation
is typically measured by the squared error $||\mathbf{Z}-\mathbf{AH}||_{F}^{2}$,
where $||\cdot||_{F}$ denotes the Frobenius norm \citep{yaghoobi2009dictionary}.
Additional assumptions on the structure of the matrices $\mathbf{A}$
and $\mathbf{H}$ can be applied in matrix factorization depending
on the application. Examples include assumptions of orthonormality
of the columns of the dictionary $\mathbf{A}$, sparsity of the coefficient
matrix $\mathbf{H}$ \citep{ramirez2012mdl}, and nonnegativity on
$\mathbf{A}$ and $\mathbf{H}$ \citep{lin2007nmf}.

We consider two popular matrix factorization methods: the singular value decomposition
(SVD) and nonnegative matrix factorization (NMF).

\subsection{Factorization using SVD}

\label{sub:dictionary_methods} To perform matrix factorization using
SVD, we take the singular value decomposition of the data matrix $\mathbf{Z}=\mathbf{U\Sigma}\mathbf{V}^{T}$
where $\mathbf{U}$ is the matrix of the left singular vectors, $\mathbf{\Sigma}$
is a diagonal matrix containing the singular values, and $\mathbf{V}$
is a matrix of the right singular vectors. If the size of the dictionary
$r$ is fixed and is less than $2m^{2}$, then the matrix of rank
$r$ that is closest to $\mathbf{Z}$ in terms of the Frobenius norm
is the matrix product $\mathbf{U}_{r}\Sigma_{r}\mathbf{V}_{r}^{T}$,
where $\mathbf{U}_{r}$ and $\mathbf{V}_{r}$ are matrices containing
only the first $r$ singular vectors and $\Sigma_{r}$ contains only
the first $r$ singular values~\citep{moon2000methods}. Thus for
SVD, the dictionary and coefficient matrices are $\mathbf{A}=\mathbf{U}_{r}$
and $\mathbf{H}=\Sigma_{r}\mathbf{V}_{r}^{T}$, respectively. Note
that SVD enforces orthonormality on the columns of $\mathbf{U}_{r}$.
Further details are included in Appendix~\ref{sub:dict_appendix}.

The intrinsic dimension estimated in \cite{2015arXiv150304127M} determines the number
of parameters required to accurately represent the data. It is used
to provide an initial estimate for the size of the dictionaries $r$. For
SVD, we choose $r$ to be one standard deviation above the mean intrinsic
dimension estimate, that is, $r \simeq 5$ or 6. The choice of $r$
is then further refined by a comparison of dictionaries in Section~\ref{sec:dict_compare}.
See Section~\ref{sub:cluster_input} for more on selecting the dictionary size.

\begin{figure}
\centering

\includegraphics[width=0.74\textwidth]{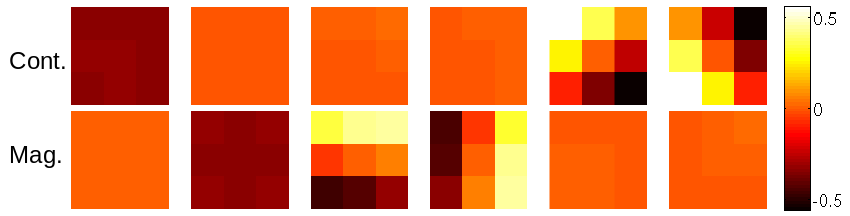}

\caption{Learned dictionary elements using SVD. Dictionary elements are constrained
to be orthonormal. The patches consist of uniform patches and gradients
in varied directions. The magnetogram patches are essentially zero
when the continuum components are nonzero and vice versa. The dictionary
size $r$ is first chosen based on the intrinsic dimension estimates
in \cite{2015arXiv150304127M} and then refined by comparing dictionaries of various sizes
in Section~\ref{sec:dict_compare}. Section~\ref{sub:cluster_input} contains more details on choosing $r$. \label{fig:pcadict}}

\end{figure}

Figure~\ref{fig:pcadict} shows the learned dictionaries using SVD
on the entire data set of 424 image pairs of ARs. Interestingly, the
SVD seems to consider the continuum and magnetogram separately as
the magnetogram elements are essentially zero when the continuum elements
are not and vice versa. This is likely caused by the orthonormality
constraint. The dictionary patches largely consist of a mix of uniform
patches and patches with gradients in varied directions. The second dictionary element is associated with the average magnetic field value of a patch.

\subsection{Factorization using NMF}

Non-negative matrix factorization (NMF) \citep{lee2001nmf} solves
the problem of minimizing $||\mathbf{Z}-\mathbf{AH}||_{F}^{2}$ while
constraining $\mathbf{A}$ and $\mathbf{H}$ to have nonnegative values.
Thus NMF is a good choice for matrix factorization when the data is
nonnegative. For our problem, the continuum data is nonnegative while
the magnetogram data is not. Therefore we use a modified version of
NMF using projected gradient where we only constrain the parts of
\textbf{$\mathbf{A}$} corresponding to the continuum to be nonnegative.
An effect of using this modified version of NMF is that since the
coefficient matrix $\mathbf{H}$ is still constrained to be nonnegative,
we require separate dictionary elements that are either positive or
negative in the magnetogram component. Thus we use approximately 1.5
times more dictionary elements for NMF than SVD. 

Since we apply NMF to the full data matrix $\mathbf{Z}$, this enforces a coupling between the two modalities by forcing the use of the same coefficient matrix to reconstruct the matrices $\mathbf{X}$ and $\mathbf{Y}$. This is similar to coupled NMF which has been used in applications such as hyperspectral and multispectral data fusion \citep{yokoya2012coupled}.

\begin{figure}
\centering

\includegraphics[width=1\textwidth]{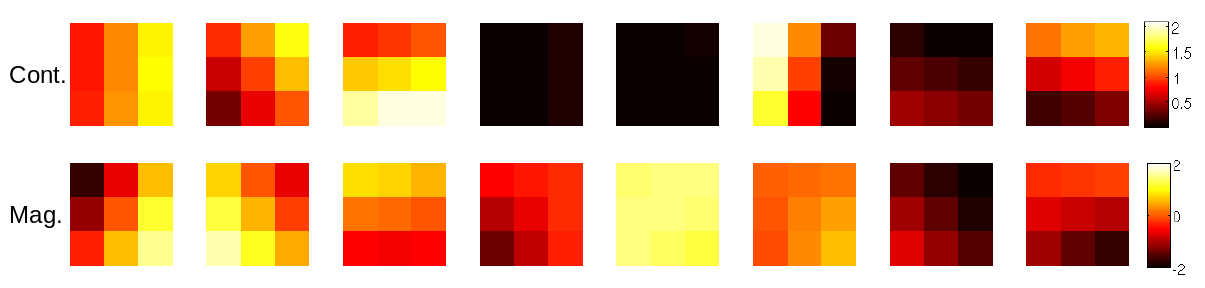}

\caption{Learned dictionary elements using NMF where the continuum dictionary
elements are constrained to be nonnegative. All the dictionary patches
consist of uniform patches or gradients in varied directions. The
order of the elements is not significant. The dictionary size $r$
is chosen to be approximately 1.5 times larger than the SVD dictionary
size, which is chosen based on the intrinsic dimension estimates in
\cite{2015arXiv150304127M} and then refined using the results in Section~\ref{sec:dict_compare}.\label{fig:nmfdict}}
\end{figure}

Figure~\ref{fig:nmfdict} shows the learned dictionary elements using
NMF on the entire dataset. For NMF, the modalities are not treated
separately as in the SVD results. But as for SVD, the patches largely
consist of a mix of uniform patches and patches with gradients in
varied directions. 

\subsection{SVD vs. NMF}

There are advantages to both SVD and NMF matrix factorization methods
which are summarized in Table~\ref{tab:SVDvsNMF}. SVD produces the
optimal rank $r$ approximation of $\mathbf{Z}$, is fast and unique,
and results in orthogonal elements \citep{langville2006nmf}. NMF
has the advantages of nonnegativity, sparsity, and interpretability.
The interpretability comes from the additive parts-based representation
inherent in NMF \citep{langville2006nmf}. In contrast, the SVD results are not sparse which can make interpretation more difficult. However, NMF is not as
robust as SVD as the NMF algorithm is a numerical approximation to
a non-convex optimization problem having local minima. Thus the solution provided by the NMF algorithm depends on the initialization. More details
on matrix factorization using NMF and SVD are included in Appendix~\ref{sub:dict_appendix}.

\begin{table}
\caption{Summary of the advantages of SVD and NMF matrix factorization methods.
The advantages of one method complement the disadvantages of the other
\citep{langville2006nmf}. For example, the NMF optimization problem is nonconvex with local minima resulting in solutions that depend on the initialization of the algorithm. \label{tab:SVDvsNMF}}
\centering

\begin{tabular}{|l|l|}
\hline 
SVD Advantages & NMF Advantages\tabularnewline
\hline
\hline 
Optimal rank $r$ approximation & Results are nonnegative\tabularnewline
\hline 
Fast to compute & Results are sparse\tabularnewline
\hline 
Unique & Sparsity and nonnegativity lead to improved interpretability\tabularnewline
\hline
\end{tabular}

\end{table}

\subsection{Methods for Comparing Matrix Factorization Results}

\label{sub:dissim}To compare the results from matrix factorization,
we primarily seek a difference between the coefficients from $\mathbf{H}$. To aid us in choosing a dictionary size $r$, we also require a measure of difference between dictionaries $\mathbf{A}$. We use the Hellinger distance and Grassmannian projection metric to measure the respective differences.

In~\cite{moon2014icip}, the Frobenius norm was used to compare the
dictionaries. However, this fails to take into
account the fact that two dictionaries may have the same elements
but in a different order. In this case, the Frobenius norm of the
difference between two dictionaries may be high even though the dictionaries span the same
subspace. A better way to measure the difference would be to compare
the subspaces spanned by the dictionaries. The Grassmannian $\mathbf{Gr}(r,V)$
is a space which parameterizes all linear subspaces with dimension
$r$ of a vector space $V$. As an example, the Grassmannian $\mathbf{Gr}\left(2,\mathbb{R}^{n}\right)$
is the space of planes through the origin of the standard Euclidean
vector space in $\mathbb{R}^{n}$. In our case, we are concerned with
the Grassmannian $\mathbf{Gr}\left(r,\mathbb{R}^{18}\right)$, where
$r$ is the size of the dictionary. The space spanned by a given dictionary
$\mathbf{A}$ is then a single point in $\mathbf{Gr}\left(r,\mathbb{R}^{18}\right)$.
Several metrics have been defined on this space including the Grassmannian
projection metric \citep{stewart1973grass,edelman1998grass}. It can
be defined as \[
d_{G}(\mathbf{A},\mathbf{A}')=\left\Vert \mathbf{P_{A}}-\mathbf{P}_{\mathbf{A}'}\right\Vert ,\]
where $\mathbf{P}_{\mathbf{A}}=\mathbf{A}\left(\mathbf{A}^{T}\mathbf{A}\right)^{-1}\mathbf{A}^{T}$
is the projection matrix of $\mathbf{A}$ and $\left\Vert \cdot\right\Vert $
is the $\ell_{2}$ norm. This metric is invariant to the order of
the dictionary elements and compares the subspaces spanned by the
dictionaries. This metric has a maximum value of 1. 

To compare the coefficient matrices, we assume that the 18-dimensional
pixel patches within an AR are samples from an 18-dimensional probability
distribution, and that each AR has a corresponding (potentially unique)
probability distribution of pixel patches. We project these samples
onto a lower dimensional space by matrix factorization. In other words,
we learn a dictionary $\mathbf{A}$ and the coefficient matrix $\mathbf{H}$
for the entire dataset $\mathbf{Z}$, and then separate the coefficients
in $\mathbf{H}$ according to the $K$ different ARs (or groups of
ARs) considered: $\mathbf{Z}=\mathbf{A}\left(\begin{array}{cccc}
\mathbf{H}_{1} & \mathbf{H}_{2} & \dots & \mathbf{H}_{K}\end{array}\right)$. The columns of $\mathbf{H}_{i}$ are a collection of projected,
low-dimensionality, samples from the $i$th AR (or group), and we
let $f_{i}$ denote the corresponding probability density function.
Given two such collections, we can estimate the difference between
their probability distributions by estimating the \emph{information
divergence}. Many kinds of divergences exist such as the popular Kullback-Leibler
divergence~\citep{kullback1951div}. We use the Hellinger distance
which is defined as\textbf{ }\citep{hellinger1909,bhattacharyya1946div,csiszar1967div}
\[
H^2(f_{i},f_{j})=1-\int\sqrt{f_{i}(x)f_{j}(x)}dx,\]
where $f_{i}$ and $f_{j}$ are the two probability densities being
compared. The Hellinger distance has the advantage over other divergences
of being a metric which is not true of divergences in general. To
estimate the Hellinger distance, we use the nonparametric estimator
derived in \cite{moon2014isit,moon2014nips} that is based on the
$k$-nearest neighbor density estimators for the densities $f_{i}$
and $f_{j}$. This estimator is simple to implement and achieves the
parametric convergence rate when the densities are sufficiently smooth.

\section{Comparisons of General Matrix Factorization Results}

\label{sec:dict_compare}We apply the metrics mentioned in Section~\ref{sub:dissim}
and compare the local features as extracted by matrix factorization
per Mount Wilson class. One motivation for these comparisons is to investigate differences between the Mount Wilson classes based on the Hellinger distance. Another motivation is to further refine our choice of dictionary size $r$ in preparation for clustering the ARs. When comparing the dictionary coefficients using the Hellinger distance, we want a single, representative dictionary that is able to accurately reconstruct all of the images. Then the ARs will be differentiated based on their respective distributions of dictionary coefficients instead of the accuracy of their reconstructions. The coefficient distributions can then be compared to interpret the clustering results as is done in Section~\ref{sub:clust_Hellinger}.

Recall that our goal is to use unsupervised methods to separate the data based on the natural geometry. Our goal is not to replicate the Mount Wilson results. Instead we use the Mount Wilson labels in this section as a vehicle for interpreting the results.

\subsection{Grassmannian Metric Comparisons}

\label{sub:GrassCompare} We first learn
dictionaries for each of the Mount Wilson types by applying matrix factorization to a subset of the patches corresponding to sunspot groups of
the respective type. We then use the Grassmannian metric to compare
the dictionaries. For example, if we want to compare the $\alpha$
and $\beta$ groups, we collect a subset of patches from all ARs designated
as $\alpha$ groups into a single data matrix $\mathbf{Z}_{\alpha}$.
We then factor this matrix with the chosen method to obtain $\mathbf{Z}_{\alpha}=\mathbf{A}_{\alpha}\mathbf{H}_{\alpha}$.
Similarly, we obtain $\mathbf{Z}_{\beta}=\mathbf{A}_{\beta}\mathbf{H}_{\beta}$
and then calculate $d_{G}(\mathbf{A}_{\alpha},\mathbf{A}_{\beta})$. 

The reason we use only a subset of patches is that each AR type has a different number of total patches (see Table~\ref{tab:mwilson_classification}) which may introduce bias in the comparisons. One source of potential bias in this case is due to the potentially increased patch variability in groups with more patches, which would result in increased difficulty in characterizing certain homogeneities of the patch features. This is mitigated somewhat by the fact that the local intrinsic dimension is typically less than 6 \citep{2015arXiv150304127M}. However, it is possible that there may be different local subspaces with the same dimension. A second source of potential bias is in the different levels of variance of the estimates due to difference in patch numbers. To circumvent these potential biases, we use the same number of patches in each group for each comparison. For the inter-class comparison, we randomly take 13,358 patches (the number of patches in the smallest class) from each class to learn the dictionary, and then calculate the Grassmannian metric. For the intra-class comparison, we take two disjoint subsets of 6,679 patches (half the number of patches in the smallest class) from each class to learn the dictionaries. This process is repeated 100 times and the resulting mean and standard deviation are reported.

\begin{table}

\caption{Difference between dictionaries learned from the collection of sunspot
patches corresponding to the different Mount Wilson types as measured
by the Grassmannian metric $d_{G}$, e.g. $d_{G}(\mathbf{A}_{\alpha},\mathbf{A}_{\beta})$. Dictionaries are learned using random subsets of the data and the results are reported in the form of mean$\pm$standard deviation using 100 trials.
Different sizes of dictionaries $r$ are used. The SVD results are sensitive to $r$. \label{tab:mwilsonGrass}}
\centering
\begin{tabular}{|c|c|c|c|c|}
\multicolumn{1}{c}{} & \multicolumn{4}{c}{SVD, Pooled Grassmannian, $r=5$}\tabularnewline
\cline{2-5} 
\multicolumn{1}{c|}{} & $\alpha$ & $\beta$ & $\beta\gamma$ & $\beta\gamma\delta$ \tabularnewline
\hline 
$\alpha$ & $0.15\pm0.10$ &  $0.26\pm0.18$ &  $0.89\pm0.06$ &  $0.34\pm0.18$  \tabularnewline
\hline 
$\beta$  &  & $0.50\pm0.29$ &  $0.89\pm0.14$ &  $0.43\pm0.27$  \tabularnewline
\hline 
$\beta\gamma$ & &  & $0.24\pm0.16$ & $0.7\pm0.2$ \tabularnewline
\hline
$\beta\gamma\delta$ & & & & $0.45\pm0.28$ \tabularnewline
\hline
\end{tabular} 

\begin{tabular}{|c|c|c|c|c|}
\multicolumn{1}{c}{} & \multicolumn{4}{c}{SVD, Pooled Grassmannian, $r=6$}\tabularnewline
\cline{2-5} 
\multicolumn{1}{c|}{} & $\alpha$ & $\beta$ & $\beta\gamma$ & $\beta\gamma\delta$ \tabularnewline
\hline 
$\alpha$ & $0.02\pm0.004$ &  $0.03\pm0.003$ &  $0.02\pm0.004$ &  $0.04\pm0.005$  \tabularnewline
\hline 
$\beta$  &  & $0.02\pm0.005$ &  $0.02\pm0.004$ &  $0.03\pm0.006$  \tabularnewline
\hline 
$\beta\gamma$ & &  & $0.03\pm0.006$ & $0.03\pm0.006$ \tabularnewline
\hline
$\beta\gamma\delta$ & & & & $0.03\pm0.007$ \tabularnewline
\hline
\end{tabular} 

\begin{tabular}{|c|c|c|c|c|}
\multicolumn{1}{c}{} & \multicolumn{4}{c}{NMF, Pooled Grassmannian, $r=8$}\tabularnewline
\cline{2-5} 
\multicolumn{1}{c|}{} & $\alpha$ & $\beta$ & $\beta\gamma$ & $\beta\gamma\delta$ \tabularnewline
\hline 
$\alpha$ & $0.40\pm0.13$ &  $0.40\pm0.10$ &  $0.33\pm0.09$ &  $0.37\pm0.10$  \tabularnewline
\hline 
$\beta$  &  & $0.29\pm0.13$ &  $0.35\pm0.09$ &  $0.37\pm0.11$  \tabularnewline
\hline 
$\beta\gamma$ & &  & $0.37\pm0.12$ & $0.34\pm0.10$ \tabularnewline
\hline
$\beta\gamma\delta$ & & & & $0.41\pm0.11$ \tabularnewline
\hline
\end{tabular} 

\begin{tabular}{|c|c|c|c|c|}
\multicolumn{1}{c}{} & \multicolumn{4}{c}{NMF, Pooled Grassmannian, $r=9$}\tabularnewline
\cline{2-5} 
\multicolumn{1}{c|}{} & $\alpha$ & $\beta$ & $\beta\gamma$ & $\beta\gamma\delta$ \tabularnewline
\hline 
$\alpha$ & $0.62\pm0.25$ &  $0.41\pm0.15$ &  $0.45\pm0.19$ &  $0.40\pm0.13$  \tabularnewline
\hline 
$\beta$  &  & $0.54\pm0.23$ &  $0.49\pm0.19$ &  $0.44\pm0.19$  \tabularnewline
\hline 
$\beta\gamma$ & &  & $0.53\pm0.23$ & $0.44\pm0.20$ \tabularnewline
\hline
$\beta\gamma\delta$ & & & & $0.49\pm0.20$ \tabularnewline
\hline
\end{tabular}

\end{table}

Table~\ref{tab:mwilsonGrass} shows the corresponding average Grassmannian
distance metrics when using SVD and NMF and for different sizes of
dictionaries $r$. For SVD, the results are very sensitive to $r$.
Choosing $r=5$ results in large differences between the different
dictionaries but for $r=6$, the dictionaries are very similar. This
suggests that for SVD, 6 principal components are sufficient to accurately
represent the subspace upon which the sunspot patches lie. This is
consistent with the results of \cite{2015arXiv150304127M} where the intrinsic
dimension is found to be less than 6 for most patches. 

Interestingly, for the $r=5$ SVD results, the $\beta\gamma$ group is the most dissimilar to the other groups while being relatively similar to itself. In contrast, the $\beta$ group is fairly dissimilar to itself and relatively similar to the $\alpha$ and $\beta\gamma\delta$ groups.

The NMF results are less sensitive to $r$. The average difference between
the dictionaries and its standard deviation is larger when $r=9$ compared to when $r=8$. However, for a given $r$, all of the mean differences are within a standard deviation of each other. Thus on aggregate, the NMF dictionaries learned from large collections of patches from multiple images differ from each other to the same degree regardless of the Mount Wilson type.

\subsection{Hellinger Distance Comparisons}

\label{sub:HellingerCompare}For the Hellinger distance, we learn
a dictionary $\mathbf{A}$ and the coefficient matrix $\mathbf{H}$
for the entire data set $\mathbf{Z}$. We then separate the coefficients
in $\mathbf{H}$ according to the Mount Wilson type and compare
the coefficients' distributions using the Hellinger distance. For
example, suppose that the data matrix is arranged as $\mathbf{Z}=\left(\begin{array}{cccc}
\mathbf{Z}_{\alpha} & \mathbf{Z}_{\beta} & \mathbf{Z}_{\beta\gamma} & \mathbf{Z}_{\beta\gamma\delta}\end{array}\right)$. This is factored as $\mathbf{Z}=\mathbf{A}\left(\begin{array}{cccc}
\mathbf{H}_{\alpha} & \mathbf{H}_{\beta} & \mathbf{H}_{\beta\gamma} & \mathbf{H}_{\beta\gamma\delta}\end{array}\right)$. To compare the $\alpha$ and $\beta$ groups, we assume that
the columns in $\mathbf{H}_{\alpha}$ are samples from the distribution
$f_{\alpha}$ and similarly $\mathbf{H}_{\beta}$ contains samples
from the distribution $f_{\beta}$. We then estimate the Hellinger distance $H(f_{\alpha},f_{\beta})$
using the divergence estimator in \cite{moon2014isit}.

When the Hellinger distance is used to compare the collections of
dictionary coefficients within the sunspots, the groups are very similar,
especially when using SVD (Table~\ref{tab:mwilsonHell}). This indicates
that when the coefficients of all ARs from one class are grouped together,
the distribution looks similar to the distribution of the other classes.
However, there are some small differences. First the intraclass distances are often much smaller than the interclass distances which indicates that there is some relative difference between most classes. Second, for both matrix factorization
methods, the $\beta\gamma\delta$ groups are the most dissimilar.
This could be due to the presence of a $\delta$ spot configuration,
where umbrae of opposite polarities are within a single penumbra.
Such a configuration may require specific linear combinations of the dictionary elements as compared
to the other classes. The presence and absence of these linear combinations in two Mount Wilson types would result in a higher Hellinger distance between them.

\begin{table}

\caption{Difference between the collection of dictionary coefficients pooled
from the different Mount Wilson classes as measured by the Hellinger
distance. Intraclass distances are reported in the form of mean$\pm$standard deviation and are calculated by randomly splitting the data and then calculating the distance over 100 trials. The size of the dictionaries is $r=6$ and $8$ for SVD
and NMF, respectively. The $\beta\gamma\delta$ group is most dissimilar
to the others.\label{tab:mwilsonHell}}
\centering

\begin{tabular}{|c|c|c|c|c|}
\multicolumn{1}{c}{} & \multicolumn{4}{c}{SVD, Pooled Hellinger}\tabularnewline
\cline{2-5} 
\multicolumn{1}{c|}{} & $\alpha$ & $\beta$ & $\beta\gamma$ & $\beta\gamma\delta$\tabularnewline
\hline 
$\alpha$ & $0.0006\pm0.004$ & 0 & 0 & 0.03\tabularnewline
\hline 
$\beta$ & & $0.0005\pm0.002$ & 0.01 & 0.08\tabularnewline
\hline 
$\beta\gamma$ & &  & $0.0003\pm0.002$ & 0.05\tabularnewline
\hline
$\beta\gamma\delta$ &  &  &  & $0.0004\pm0.002$\tabularnewline
\hline
\end{tabular} \begin{tabular}{|c|c|c|c|c|}
\multicolumn{1}{c}{} & \multicolumn{4}{c}{NMF, Pooled Hellinger}\tabularnewline
\cline{2-5} 
\multicolumn{1}{c|}{} & $\alpha$ & $\beta$ & $\beta\gamma$ & $\beta\gamma\delta$\tabularnewline
\hline 
$\alpha$ & $0\pm0$ & 0.08 & 0.05 & 0.10\tabularnewline
\hline 
$\beta$ & & $0.00007\pm0.0004$ & 0.03 & 0.12\tabularnewline
\hline 
$\beta\gamma$ & &  & $0.000002\pm0.00003$  & 0.11\tabularnewline
\hline
$\beta\gamma\delta$ &  &  &  & $0.00001\pm0.0002$\tabularnewline
\hline
\end{tabular}

\end{table}

Again, for clustering, we compute the pairwise Hellinger distance
between each AR's collection of coefficients. This is done by forming
the data matrix from the 424 ARs as $\mathbf{Z}=\left(\begin{array}{cccc}
\mathbf{Z}_{1} & \mathbf{Z}_{2} & \dots & \mathbf{Z}_{424}\end{array}\right)$ and factoring it as $\mathbf{Z}=\mathbf{A}\left(\begin{array}{cccc}
\mathbf{H}_{1} & \mathbf{H}_{2} & \dots & \mathbf{H}_{424}\end{array}\right)$. The columns of $\mathbf{H}_{i}$ are samples from a distribution
$f_{i}$ and the distributions $f_{i}$ and $f_{j}$ are compared
by estimating $H(f_{i},f_{j})$. The corresponding dictionaries for the two methods are shown in Figures~\ref{fig:pcadict} and~\ref{fig:nmfdict}.

Table~\ref{tab:avgHell} gives the average pairwise Hellinger distance between
the ARs. The average distances differ more with the NMF based coefficients
resulting in larger dissimilarity. The average distance is smallest
when comparing the $\beta$ groups to all others and largest when
comparing the $\beta\gamma$ groups to the rest. The standard deviation is also larger when comparing $\alpha$ and $\beta$ groups. This may be partially related to the variability in estimation due to smaller sample sizes as the $\alpha$ and $\beta$ groups contain more of the smaller ARs (see Figure~\ref{fig:patch}).  Overall,
the average distances show that there are clear differences between
the ARs within the sunspots using this metric.

\begin{table}
\caption{Average pairwise difference between dictionary coefficients from each
AR from different Mount Wilson types as measured by the Hellinger
distance. Results are reported in the form of mean$\pm$standard deviation. The size of the dictionaries is $r=6$ and $8$ for SVD
and NMF, respectively. The $\beta\gamma$ ARs are most dissimilar
to each other and the other classes while the $\beta$ ARs are most
similar.\label{tab:avgHell}}
\centering

\begin{tabular}{|c|c|c|c|c|}
\multicolumn{1}{c}{} & \multicolumn{4}{c}{SVD, Average Hellinger}\tabularnewline
\cline{2-5} 
\multicolumn{1}{c|}{} & $\alpha$ & $\beta$ & $\beta\gamma$ & $\beta\gamma\delta$\tabularnewline
\hline 
$\alpha$ & $0.83\pm0.21$ & $0.80\pm0.20$ & $0.82\pm0.16$ & $0.80\pm0.14$ \tabularnewline
\hline 
$\beta$ &  & $0.75\pm0.22$ & $0.78\pm0.18$ & $0.77\pm0.17$ \tabularnewline
\hline 
$\beta\gamma$ &  &  & $0.83\pm0.15$ & $0.81\pm0.14$\tabularnewline
\hline 
$\beta\gamma\delta$ &  &  &  & $0.78\pm0.13$\tabularnewline
\hline
\end{tabular} \begin{tabular}{|c|c|c|c|c|}
\multicolumn{1}{c}{} & \multicolumn{4}{c}{NMF, Average Hellinger}\tabularnewline
\cline{2-5} 
\multicolumn{1}{c|}{} & $\alpha$ & $\beta$ & $\beta\gamma$ & $\beta\gamma\delta$\tabularnewline
\hline 
$\alpha$ & $0.85\pm0.21$ & $0.81\pm0.20$ & $0.84\pm0.17$ & $0.82\pm0.14$\tabularnewline
\hline 
$\beta$ &  & $0.76\pm0.23$ & $0.80\pm0.19$ & $0.80\pm0.17$\tabularnewline
\hline 
$\beta\gamma$ &  &  & $0.85\pm0.15$ & $0.85\pm0.14$\tabularnewline
\hline 
$\beta\gamma\delta$ &  &  &  & $0.83\pm0.12$ \tabularnewline
\hline
\end{tabular}
\end{table}

\section{Clustering of Active Regions: Methods}

\label{sec:cluster}

\subsection{Clustering Algorithm}

The clustering algorithm we use is the Evidence Accumulating Clustering
with Dual rooted Prim tree Cuts (EAC-DC) method in~\cite{galluccio2013clustering}
which scales well for clustering in high dimensions. EAC-DC clusters
the data by defining a metric based on the growth of two minimal spanning
trees (MST) grown sequentially from a pair of points. To grow the
MSTs, a base dissimilarity measure is required as input such as the Hellinger distance described
in Section~\ref{sub:dissim}. From the new metric defined using the
MSTs, a similarity measure between inputs is created. It is fed into
a spectral clustering algorithm that groups together inputs which
are most similar. The similarity measure based on the MSTs adapts
to the geometry of the data, and this results in a clustering method
that is robust and competitive with other algorithms \citep{galluccio2013clustering}.
See Appendix~\ref{sub:EACDC} for more details.

\subsection{Clustering Input: Dictionary Sizes}

\label{sub:cluster_input}As input to the clustering algorithm, we
use the matrix factorization results as described in Section~\ref{sub:dissim}.
We learn a single dictionary from
the entire dataset. We then project the data onto a lower dimensional
space, i.e. we learn the coefficient matrices \textbf{$\mathbf{H}_{i}$}. These matrices are the inputs in this method and the base dissimilarity measure is the Hellinger distance estimated using each AR's respective coefficients.
 Table~\ref{tab:measures}
provides a summary of the various dissimilarity and similarity measures
that we use.

\begin{table}

\caption{Summary of the dissimilarity and similarity measures used.\label{tab:measures}}
\centering

\begin{tabular}{|l|l|l|}
\hline 
Measure & Type & Properties\tabularnewline
\hline
\hline 
Grassmannian metric & Dissimilarity & Compares dictionaries by comparing the subspace \tabularnewline
 &  & spanned by the dictionary elements\tabularnewline
\hline 
Hellinger distance & Dissimilarity & Compares the underlying distributions of dictionary \tabularnewline
 &  & coefficients; estimated using \cite{moon2014isit}\tabularnewline
\hline 
EAC-DC based measure & Similarity & Based on sequentially grown MSTs of the data; \tabularnewline
 &  & requires a base dissimilarity measure as input\tabularnewline
\hline
\end{tabular}

\end{table}

As mentioned in Section~\ref{sec:dictionary}, the estimated intrinsic
dimension from \cite{2015arXiv150304127M} is used to provide an initial estimate for the size
of the dictionaries $r$. The choice of $r$ is further refined by
the dictionary comparison results from Section~\ref{sec:dict_compare}
. For SVD, we choose $r$ to be one standard deviation above the mean
intrinsic dimension estimate which is approximately 5 or 6. When
comparing the dictionary coefficients, we want the single dictionary
to accurately represent the images. The dictionary
should not be too large as this may add spurious dictionary elements
due to the noise. The results in Table~\ref{tab:mwilsonGrass} suggest
that for SVD, the dictionaries are essentially identical for $r=6.$
This means that $6$ dictionary elements are sufficient to accurately
reconstruct most of the images. This is consistent with the intrinsic
dimension estimates in \cite{2015arXiv150304127M}. Thus we choose $r=6$ when using the
Hellinger distance for the SVD dictionary coefficients. 

Since our mixed version of NMF requires approximately 1.5 times the
number of dictionary elements as SVD (see Section~\ref{sub:dictionary_methods}),
we choose $r=8$ within the sunspots. Since the differences between classes were similar for $r=8$ and $r=9$, choosing $r=8$ strikes a balance between
accurate representation of the data
and limiting the effects of noise.

\subsection{Clustering Input: Patches within Sunspots and Along the Neutral Line}

\textbf{\label{sub:neutralline_method}} Our main focus up to this point in this paper
has been on data matrices $\mathbf{Z}$ containing the patches within the
STARA masks, that is, within sunspots. The clustering based on these
patches is discussed in Sections~\ref{sub:clust_Hellinger}-\ref{sub:discussion}.

It is well-known, however, that the shape of the neutral line separating
the main polarities plays an important role in the Mount Wilson classification.
For this reason, we conduct two experiments involving data from along the neutral line. 

The results of the first experiment are in Section~\ref{sub:neutral} where we apply matrix factorization on a data matrix containing only
the patches situated along the neutral line using the same ARs as in Sections ~\ref{sub:clust_Hellinger}-\ref{sub:discussion}. To compute the location of a neutral line in this experiment, we assume it is situated
in the middle between regions of opposite polarity, and proceed as
follows. First, we determine regions of high magnetic flux of each
polarity using an absolute threshold at 50 Gauss. Second, we compute
for each pixel the distance to the closest high flux region in each
polarity using the Fast Marching method \citep{Sethian95afast}.
Once the two distance fields (one for each polarity) are calculated,
the neutral line can be obtained by finding the pixels that lie on
or are close to the zero level set of the difference of these two
distance fields. In this paper, we choose a maximum distance of 10
pixels to determine the neutral line region. 

We extract the patches that lie in the neutral line region and within
the SMART mask associated to the AR. Call the resulting data matrix $\mathbf{Z}_{N}$. We then apply SVD or NMF matrix factorization as before and calculate the pairwise distance between each
AR neutral line using the Hellinger distance on the results. Define the resulting $424\times 424$  dissimilarity matrix as $\mathbf{D}_{N}$ for whichever factorization method we are currently using. Similarly, define $\mathbf{Z}_{S}$ and $\mathbf{D}_{S}$ as the respective data matrix and dissimilarity matrix of the data from within the sunspots using the same configuration. The base dissimilarity measure $\mathbf{D}$ inputted in
the clustering algorithm is now a \emph{weighted average} of the distances computed within the neutral line regions and
within the sunspots: $\mathbf{D}=w\mathbf{D}_{N}+(1-w)\mathbf{D}_S$ where $0\leq w\leq 1$. Using a variety of weights, we then compare the
clustering output to different labeling schemes based on the Mount
Wilson labels as shown in Table~\ref{tab:labels}. 

\begin{table}

\caption{The labels used to compare with the clustering results when analyzing
the effects of including the neutral line.\label{tab:labels}}

\centering

\begin{tabular}{|c|c|}
\hline 
\# of clusters & Mount Wilson Comparison\tabularnewline
\hline
\hline 
2 & Simple ($\alpha$ and $\beta$), complex ($\beta\gamma$ and $\beta\gamma\delta$)\tabularnewline
\hline 
3 & $\alpha$, $\beta$, and complex\tabularnewline
\hline 
4 & Mount Wilson ($\alpha$, $\beta$, $\beta\gamma$, and $\beta\gamma\delta$)\tabularnewline
\hline
\end{tabular}

\end{table}

For the second experiment, we perform clustering on a ROI that selects pixels along a strong field polarity reversal line. The high gradients near strong field polarity reversal lines in LOS magnetograms are a proxy for the presence of near-photospheric electrical currents, and thus might be indicative of a non-potential configuration \citep{2007ApJ...655L.117S}. To compute this ROI, the magnetograms are first reprojected using an equal-area, sinusoidal re-projection that uses Singular-Value Padding \citep{deforest2004}. The latter is known to be more accurate than image interpolation on transformed coordinates. To conserve flux, the magnetograms are also area-normalized.

In the reprojected magnetograms, we delimit an AR using sunspot information from the Debrecen catalog \citep{gyori2010photospheric} to obtain the location of all the pixels that belong to the spots related to an AR (called "Debrecen spots" hereafter). The ROI consists of a binary array constructed as follows:
\begin{enumerate}
\item The pixels that belong to the Debrecen spots are assigned the scalar value 1 and all others are assigned the scalar value -1. 
\item From this two-valued array, we retrieve the distances from the zero level-set using a fast marching method \citep{Sethian95afast} implemented in the Python SciKits' module "scikit-fmm". 
\item We mask out the pixels within a distance of 80 pixels from the zero level-set  (in the equal-area-reprojected coordinate system). 
\item The ROI is delimited by the convex hull of the resulting mask, that is, the smallest convex polygon that surrounds all 1-valued pixels. The resulting mask is a binary array with the pixels inside the convex hull set to True. 
\end{enumerate}
Within the ROI, the $R$-value is calculated similarly to the method described in \cite{2007ApJ...655L.117S}. We dilate the image using a dilation factor of 3 pixels, and extract the flux using overlapping Gaussian masks with $\sigma = 2\,px$. Integrating the non-zero values outputs the $R$-value, i.e, the total flux in the vicinity of the polarity-inversion line.

We then perform an image patch
analysis using image patches from this region. We do this by using either SVD or NMF to do dimensionality reduction on the patches, and then estimate
the Hellinger distance between ARs using the reduced dimension representation. We exclude $\alpha$ groups
from the analysis as they do not have a strong field polarity reversal line. This leaves 420
images to be clustered. The clustering assignments are then compared to the calculated $R$ value via the correlation coefficient. The results are presented in Section~\ref{sub:rvalue}.

\section{Clustering of Active Regions: Results}

\label{sec:cluster_results}Given the choices of matrix factorization
techniques (NMF and SVD)  we have two different clustering
results on the data. Section~\ref{sub:clust_Hellinger} focuses on the clusterings using data from within the sunspots, and Section~\ref{sub:discussion} provides some recommendations for which metrics and matrix factorization techniques to use to study different ARs. The neutral line clustering results are then given in Section~\ref{sub:neutral} followed by the $R$ value based experiment in Section~\ref{sub:rvalue}.

\subsection{Clustering within the Sunspot}

\label{sub:clust_Hellinger} We now present the clustering results when using the Hellinger distance as the base dissimilarity. The corresponding dictionary elements to the coefficients are represented in Figure~\ref{fig:pcadict}
(for the SVD factorization) and in Figure~\ref{fig:nmfdict} (for
the NMF).

The EAC-DC algorithm does not automatically choose the number of clusters.
We use the mean silhouette heuristic to determine the most natural
number of clusters as a function of the data \citep{rousseeuw1987silhouettes}.
The silhouette is a measure of how well a data point belongs to its
assigned cluster. The heuristic chooses the number of clusters that
results in the maximum mean silhouette. In both clustering
configurations, the number of clusters that maximizes the mean silhouette
is 2 so we focus on the two clustering case for all clustering
schemes throughout.

To compare the clustering correspondence, we use the adjusted
Rand index (ARI). The ARI is a measure of similarity between two clusterings
(or a clustering and labels) and takes values between -1 and 1. A
1 indicates perfect agreement between the clusterings and a 0 corresponds
to the agreement from a random assignment \citep{rand1971index}.
Thus a positive ARI indicates that the clustering correspondence is better than a
random clustering while a negative ARI indicates it is worse. The ARI between the NMF and SVD clusterings is $0.27$ which indicates some overlap.

 Visualizing the clusters in lower dimensions is done with multidimensional
scaling (MDS) as in~\cite{moon2014icip}.
Let  $\mathbf{S}$ be the $424\times424$ symmetric matrix that contains the AR pair similarities as created by  EAC-DC algorithm. MDS projects the similarity matrix $\mathbf{S}$  onto the eigenvectors of the
normalized Laplacian of $\mathbf{S}$ \citep{kruskal1978mds}.
Let $\mathbf{c}_i\in\mathbb{R}^{424}$ be the projection onto the $i$th eigenvector of $\mathbf{S}$ using NMF. 
The first eigenvector represents the direction of highest variability in the matrix $\mathbf{S}$, and hence a high value of the $k-$th element of $\mathbf{c}_{1}$ indicates that the $k-$th AR is dissimilar to  other ARs.

\begin{figure}
\centering

\includegraphics[width=0.5\textwidth]{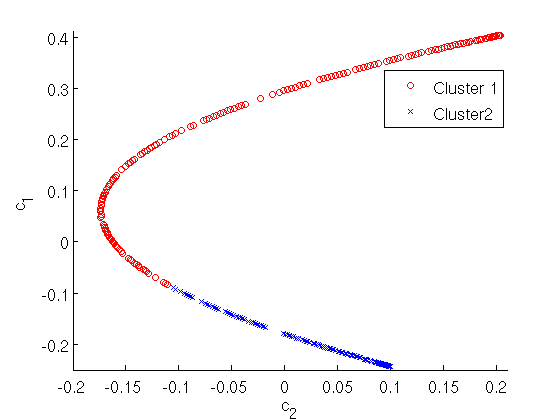}\includegraphics[width=0.5\textwidth]{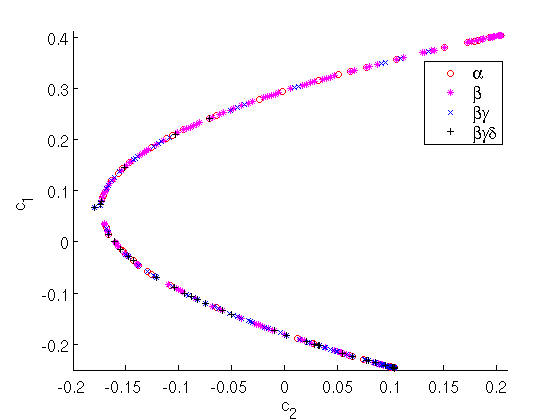}

\includegraphics[width=0.5\textwidth]{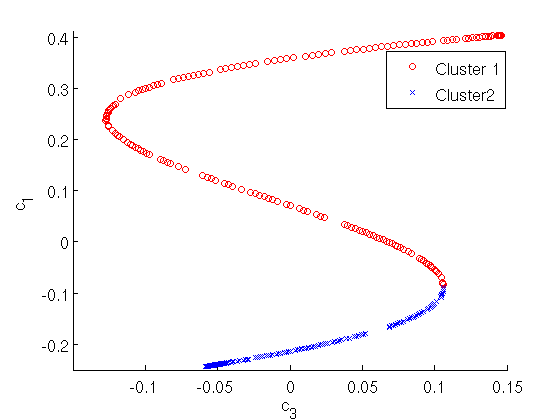}\includegraphics[width=0.5\textwidth]{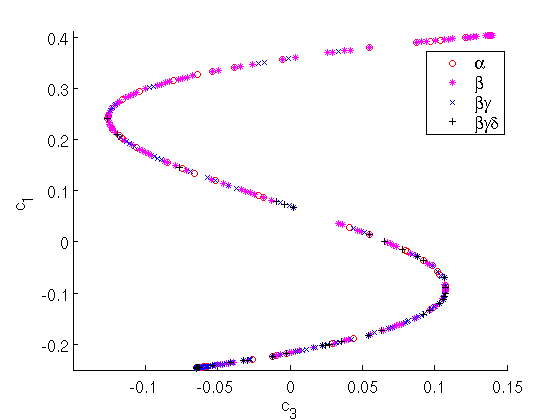}

\caption{Scatter plot of MDS variables $\mathbf{c}_{1}$ vs. $\mathbf{c}_{2}$ (top) and $\mathbf{c}_{1}$
vs. $\mathbf{c}_{3}$ (bottom) where $\mathbf{c}_{i}\in\mathbb{R}^424$ is the projection of the similarity
matrix onto the $i$th eigenvector of the normalized Laplacian of
the similarity matrix when using the NMF coefficients. Each point corresponds to one AR and they are labeled according to the clustering
(left) and the Mount Wilson labels (right). In this space, the clusters
data appear to be separable and there are concentrations of complex
ARs in the region with lowest $\mathbf{c}_{1}$ values. Other patterns are
described in the text.\label{fig:MDS}}

\end{figure}

\begin{figure}
\centering

\includegraphics[width=0.5\textwidth]{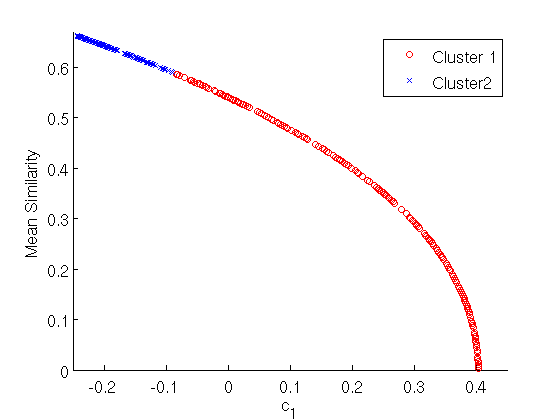}

\caption{Mean similarity of an AR with all other ARs as a function of its MDS
variable $c_{1}$ using NMF. Cluster 2 is
associated with those ARs that are most similar to all other ARs while
Cluster 1 contains those that are least similar to all others.\label{fig:Meansim} }

\end{figure}

 Figure~\ref{fig:MDS} displays the scatter plot
of $\mathbf{c}_{1}$ vs. $\mathbf{c}_{2}$ (top) and $\mathbf{c}_{1}$ vs. $\mathbf{c}_{3}$ (bottom) using NMF.  Comparing them with the Mount Wilson classification 
we see a concentration of simple ARs in the region with highest 
$\mathbf{c}_{1}$ values (most dissimilar ARs), and a concentration of   complex ARs in the region with lowest $\mathbf{c}_{1}$ (more similar ARs). 
We  can show this more precisely by computing the mean similarity of the
$i$th AR to all other ARs as the mean of the $i$th row (or column) of $\mathbf{S}$.  The value from $\mathbf{c}_{1}$  is then inversely related to this mean similarity  as seen in Figure~\ref{fig:Meansim}. 

The similarity defined under this clustering scheme gathers in Cluster 2 \lq similar' AR that are for a large part of the type $\beta\gamma$ and $\beta\gamma\delta$, whereas Cluster 1 contains AR that are more \lq dissimilar' to each other, with a large part of $\alpha$ or $\beta$ active regions.  The other clustering configuration
has a similar relationship between the first MDS coefficient and the
mean similarity.

Table~\ref{tab:Meansim} makes this clearer by showing the mean similarity
measure within each cluster and between the two clusters, which is calculated in the following manner. Suppose that the similarity
matrix is organized in block form where the upper left block corresponds
to Cluster 1 and the lower right block corresponds to Cluster 2. The
mean similarity of Cluster 1 is calculated by taking the mean of all
the values in the upper left block of this reorganized similarity
matrix. The mean similarity of Cluster 2 is found similarly from the
lower right block and the mean similarity between the clusters is
found from either the lower left or upper right blocks. These means
show that under the NMF clustering scheme, ARs in Cluster
2 are very similar to each other while ARs in Cluster 1 are not very
similar to each other on average. In fact, the ARs in Cluster 1 are
more similar to the ARs in Cluster 2 on average than to each other.
The other clustering configuration has a similar relationship between
cluster assignment and mean similarity.

%
\begin{table}
\caption{Mean similarity of ARs to other ARs either in the same cluster (1
vs. 1 or 2 vs. 2) or in the other cluster (1 vs. 2) under the different
schemes.  Cluster 1 contains ARs that
are very dissimilar to each other while Cluster 2 contains ARs that are
very similar to each other.  \label{tab:Meansim}}

\centering

\begin{tabular}{|c|c|c|c|}
\multicolumn{1}{c}{} & \multicolumn{3}{c}{Mean Similarity}\tabularnewline
\cline{2-4} 
\multicolumn{1}{c|}{} & 1 vs. 1 & 1 vs. 2 & 2 vs. 2\tabularnewline
\hline 
SVD, Hellinger & 0.29 & 0.42 & 0.87\tabularnewline
\hline 
NMF, Hellinger & 0.30 & 0.42 & 0.88\tabularnewline
\hline
\end{tabular}
\end{table}

\begin{figure}
\centering

\includegraphics[width=0.5\textwidth]{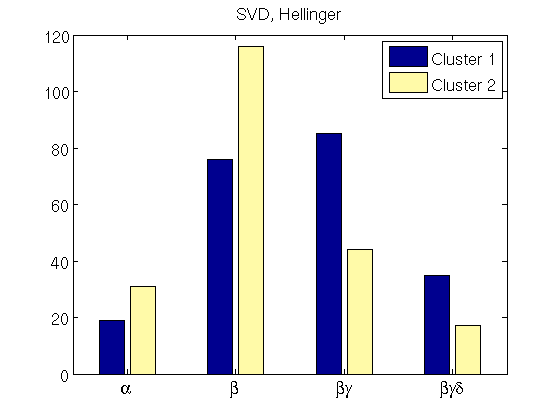}\includegraphics[width=0.5\textwidth]{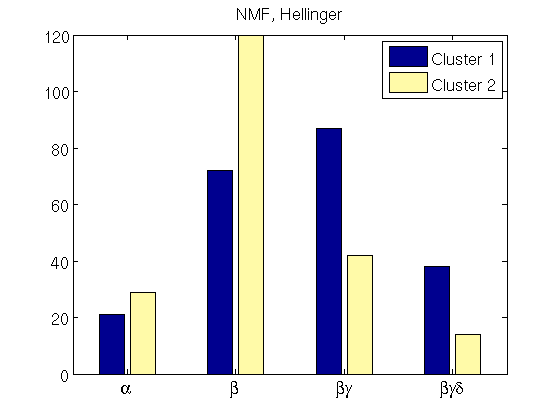}

\caption{Histograms of the Mount Wilson classes divided by clustering assignment
using the Hellinger distance.
Cluster 1 contains more of the complex ARs while Cluster 2 contains
more of the simple ARs.\label{fig:sunspothistHell}}

\end{figure}

This relationship between AR complexity and clustering assignment is further noticeable in Figure~\ref{fig:sunspothistHell} which gives a histogram of the Mount Wilson classes divided by clustering assignment. This figure shows clear patterns between
the clusterings and Mount Wilson type distribution, where the clustering separates somewhat the complex sunspots from the simple sunspots. This suggests that these configurations are clustering based on some measure of AR complexity. 

%
\begin{table}
\caption{Mean and median number of pixels of the ARs in each cluster under
the Hellinger clustering schemes. Cluster 1 contains the larger sunspots
for all groups when using NMF and for some of the groups when using
SVD. \label{tab:numspots_Hellinger}}
\centering

\begin{tabular}{|l|c|c|c|c|c|c|c|c|c|c|}
\multicolumn{1}{c}{} & \multicolumn{10}{c}{Number of Pixels}\tabularnewline
\cline{2-11} 
\multicolumn{1}{l|}{} & \multicolumn{2}{c|}{$\alpha$} & \multicolumn{2}{c|}{$\beta$} & \multicolumn{2}{c|}{$\beta\gamma$} & \multicolumn{2}{c|}{$\beta\gamma\delta$} & \multicolumn{2}{c|}{All}\tabularnewline
\hline 
Cluster & 1 & 2 & 1 & 2 & 1 & 2 & 1 & 2 & 1 & 2\tabularnewline
\hline 
Mean, SVD Hellinger & 278 & 260 & 582 & 270 & 823 & 577 & 1234 & 1354 & 756 & 422\tabularnewline
\hline 
Mean, NMF Hellinger & 384 & 183 & 788 & 156 & 847 & 515 & 1418 & 880 & 882 & 283\tabularnewline
\hline 
Median, SVD Hellinger & 148 & 128 & 477 & 70 & 612 & 472 & 1012 & 1172 & 588 & 174\tabularnewline
\hline 
Median, NMF Hellinger & 265 & 121 & 580 & 56 & 631 & 393 & 1157 & 665 & 677 & 105\tabularnewline
\hline
\end{tabular}

\end{table}

The Hellinger-based clusterings are correlated with sunspot
size for some of the Mount Wilson classes, see Table~\ref{tab:numspots_Hellinger}.
Based on the mean and median number of pixels, the Hellinger distance
on the NMF coefficients tends to gather in Cluster 2 the smallest
AR from classes $\alpha$, $\beta$, and $\beta\gamma$. Similarly,
the Hellinger distance on the SVD coefficients separates the $\beta$ and $\beta\gamma$
AR by size with Cluster 1 containing the largest and Cluster 2 containing
the smallest AR. 

Since the Hellinger distance calculates differences between ARs based on their respective distribution of dictionary coefficients, we can examine the coefficient distribution to obtain insight on what features the clustering algorithm is exploiting. For simplicity, we examine the marginal histograms of the coefficients pooled from ARs of a given cluster. When looking at the SVD coefficients, we see that their marginal distributions are similar across clusters, except for the coefficients that correspond to the second dictionary element of Figure~\ref{fig:pcadict}.
Recall that this second dictionary element is associated
with the average magnetic field value of a patch. If the corresponding
coefficient is close to zero, it means the average magnetic field
in the patch is also close to zero. 

\begin{figure}
\centering

\includegraphics[width=0.5\textwidth]{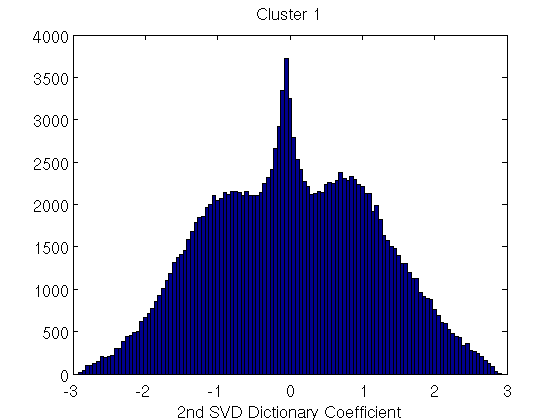}\includegraphics[width=0.5\textwidth]{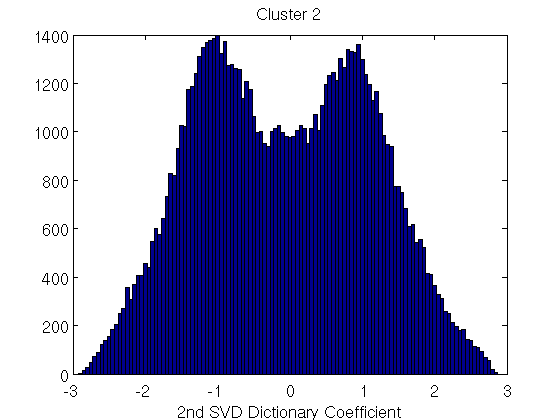}

\caption{Histograms of the marginal distributions of the coefficients corresponding
to the mean magnetic field value (dictionary element 2 in Figure~\ref{fig:pcadict})
for Cluster 1 (left) and Cluster 2 (right) using the SVD coefficients.
Cluster 1 ARs contain more patches with near neutral magnetic field
values.\label{fig:Hist_Hellinger_svd}}

\end{figure}

Figure~\ref{fig:Hist_Hellinger_svd}
shows histograms of the coefficients of the second dictionary element.
The histograms correspond to patches from all ARs separated by cluster
assignment. The histograms show that Cluster 1 has a high concentration
of patches with near zero average magnetic field. In contrast, the
larger peaks for Cluster 2 are centered around $+1$ and $-1$. This
suggests that the clustering assignments are influenced somewhat by
the amount of patches in an AR that have near zero average magnetic
field values. As we are considering only the core (sunspot) part of
the AR, having $3 \times 3$ patch with a near zero average magnetic
field entails that the corresponding patch is likely to be located
along the neutral line separating strong magnetic fields of opposite polarity. 
Thus the local distribution of magnetic field values is related to
cluster assignments when using the SVD coefficients. This is consistent with Figure~\ref{fig:sunspothistHell} where cluster 1 contains more of the complex ARs ($\beta\gamma$ and $\beta\gamma\delta$) and fewer simple ARs ($\alpha$ and $\beta$) than cluster 2 as measured by the Mt. Wilson scheme.

Checking the individual ARs and their coefficient distributions in
each cluster, we see indeed that Cluster 1 does contain more ARs with
patches having near zero average magnetic field. This tends to include
more of the complex ARs in Cluster 1 since they are more likely to
have a neutral line close to the regions of strong magnetic fields
that will therefore be included in the STARA masks.

It should be noted however that the correspondence is not perfect.
There are some ARs in Cluster 2 where the regions of opposing polarity
are close to each other and some ARs in Cluster 1 where the regions
of opposing polarity are far apart. Thus the distribution of average
magnetic field values is only one factor in the natural geometry of the ARs defined by the Hellinger distance. As mentioned
previously, the size of the AR is another factor, especially for $\beta$
groups. This is consistent with Figure~\ref{fig:sunspothistHell} where cluster 1 does contain some of the simple ARs which are less likely to have strong magnetic fields around the neutral line.

Investigating the joint histograms of the NMF dictionary elements
corresponding to positive and negative magnetic field values reveals
that the NMF Hellinger clustering results are also influenced by the
local magnetic field distribution.

All these observations indicate that the natural geometry exploited by both clustering configurations
is related to some form of complexity of the ARs.

\subsection{Discussion of Sunspot Results}

\label{sub:discussion}We note that
the Cluster 2 ARs containing the smallest sunspots are most similar
to each other while the Cluster 1 ARs are more dissimilar (see Tables~\ref{tab:Meansim} and~\ref{tab:numspots_Hellinger}). This indicates
that the Hellinger distance approaches are best for distinguishing
between different types of larger or complex ARs.

When NMF is applied on datasets where all values in the dictionary and coefficient
matrices are constrained to be nonnegative, its results are generally
more interpretable than SVD. In our application however, the magnetogram
components can be negative. Hence the NMF results are not particularly sparse
and lose some benefits of nonnegativity since the positive and negative
magnetogram components can cancel each other. This results in some
loss of interpretability. Additionally, the SVD results seem to be
more interpretable due to separate treatment of the continuum and
magnetogram components. However, there is still some value in the
NMF approach as we see that the clustering on NMF
coefficients are better at separating the ARs by size than the SVD
approach. Additionally, the NMF approaches tend to agree more strongly
with the Mount Wilson labels than their SVD counterparts as is seen in Section~\ref{sub:neutral} below.
Future work could include using alternate forms
of NMF such as in \cite{ding2010convex} where sparsity and interpretability
is preserved even when the dictionary is no longer constrained to
be nonnegative. Another variation on coupled NMF that may be applicable is soft NMF \citep{seichepine2014soft} where the requirement that the two modalities share the same regression coefficients is relaxed somewhat. Finally, future work could perform factorization using a composite objective function comprised of two terms corresponding to the two modalities that are scaled according to their noise characteristics.

\subsection{Clustering with Neutral Line Data}

\label{sub:neutral}As described in Section~\ref{sub:neutralline_method},
we analyze the effects of including data from the neutral line in
the clustering. We proceed by taking a weighted average of the dissimilarities
calculated from the sunspots and from the neutral line data matrices.
Using the ARI, we compare the results to labels based on the Mount
Wilson classification scheme, see Table~\ref{tab:labels} for the
label definition. We use a grid of weights, starting from a weight
of 0 for a clustering using only patches within sunspots up to a weight
of 1 for a clustering that takes into account only the neutral line
data.

\begin{figure}
\centering

\includegraphics[width=0.5\textwidth]{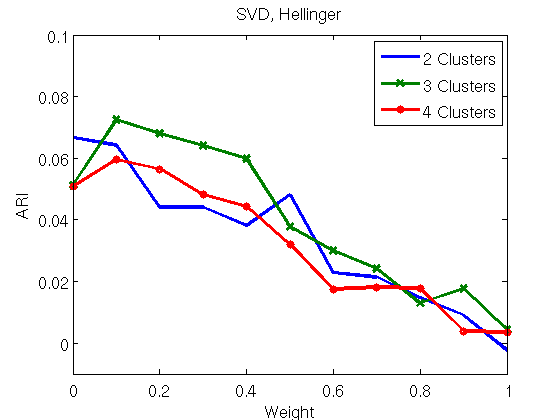}\includegraphics[width=0.5\textwidth]{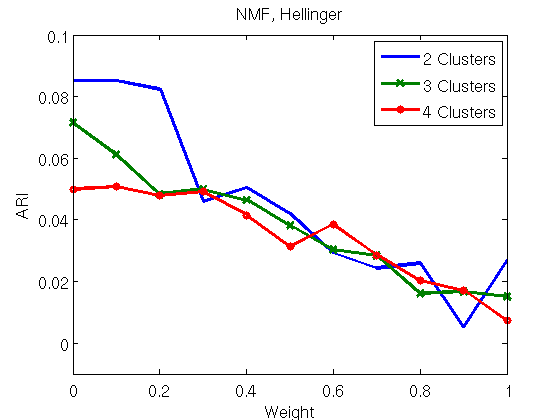}

\caption{Plot of the adjusted Rand index (ARI) using
and Hellinger distances within the neutral line and sunspots
as a function of the weight. A weight of 0 corresponds to clustering
with only the sunspots while a weight of 1 clusters with only the
neutral line. The different lines correspond to different numbers
of clusters and the corresponding labels from Table~\ref{tab:labels}.
Higher ARI indicates greater correspondence. \label{fig:ARIline}}

\end{figure}

Figure~\ref{fig:ARIline} plots the ARI for the four different schemes
as a function of the weight. In nearly all cases, the ARI is above
zero which indicates that the clustering does slightly better than
a random assignment. In general, the correspondence of the clustering
results with the Mount Wilson based labels decreases as the weight
approaches 1. This means that the natural clusterings associated with
only the neutral line data do not correspond as well with the Mount Wilson
based labels. However in several cases, including some information
from the neutral line at lower weights appears to increase the correspondence,
e.g., the ARI increases for the 3 and 4 cluster cases for the SVD coefficients. This suggests that the neutral line and the sunspots contain information about AR complexity that may be different. 


%



%

Note that clustering separates the ARs based on the natural
geometry in the spaces we are considering. Thus we can influence the
clustering by choosing the space. For example,  if we restrict our
analysis to include only coefficients corresponding to specific dictionary
patches, then this will influence the clustering.

The gradients of magnetic field values across the neutral line is
a key quantity used in several indicators of potential eruptive activity~\citep{2007ApJ...655L.117S}.
We therefore repeated the neutral line experiment where we focused
only on the gradients within the magnetogram as follows. When we applied
SVD to the data matrix $\mathbf{Z}$ extracted
from the neutral line, the resulting dictionary matrix $\mathbf{A}$
was very similar to that shown in Figure~\ref{fig:pcadict}. Note
that elements 3 and 4 correspond to the gradient patterns within the
magnetogram data. Therefore, after learning $\mathbf{A}$ and $\mathbf{H}$
from $\mathbf{Z}$, we kept only the coefficients corresponding to
dictionary elements 3 and 4, i.e. the 3rd and 4th rows of $\mathbf{H}$.
We then estimated the Hellinger distance between the ARs' underlying
distributions of these two coefficients. This restricted the neutral line analysis to include only the coefficients corresponding to magnetogram gradients. For the data within the sunspots,
we included all coefficients as before.

\begin{figure}
\centering

\includegraphics[width=0.5\textwidth]{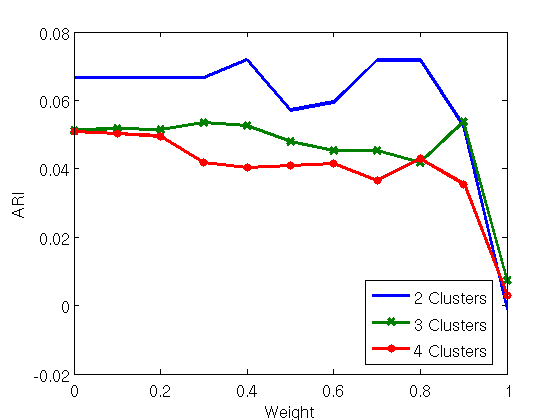}

\caption{Plot of the ARI using the Hellinger distance within the neutral line
on only the coefficients corresponding to the SVD dictionary elements
associated with magnetogram gradients. The corresponding dictionary
elements are similar to the 3rd and 4th elements in Figure~\ref{fig:pcadict}.
Focusing on the gradients results in a higher ARI for higher weights
than when all coefficients are used as seen in Figure~\ref{fig:ARIline}.
\label{fig:ARIGradient}}

\end{figure}

Figure~\ref{fig:ARIGradient} shows the ARI as a function of the
weight for this experiment. For all cases, the ARI stays fairly
constant until the weight increases
to 0.9, after which it drops dramatically. We can compare this to the results in Figure~\ref{fig:ARIline} (left) to determine if using only the neutral line gradient coefficients results in increased correspondence with the Mount Wilson labels relative to using all of the neutral line coefficients. From this comparison, the ARI is higher when using only the gradient components for weights greater than 0.1 and less than 1. Thus the correspondence
with the Mount Wilson labels and the clustering is higher when we only include the magnetogram gradient coefficients. Since the Mount Wilson scheme is related to the complexity of the neutral line, this higher correspondence suggests that focusing on the gradients in the neutral line results in a natural geometry that is more closely aligned with the complexity of the neutral line than simply using all of the coefficients. Applying supervised techniques would lead to improved correspondence.

The clear patterns in the ARI indicate
that the relationship between the weight and the ARI is unlikely to
be due entirely to noise. Thus including data from the neutral line with the data from the sunspots would add value in an unsupervised setting and would likely lead to improved performance in a supervised setting.

\subsection{Clustering of Regions Exhibiting Strong Field Polarity Reversal Lines}

\label{sub:rvalue}We now analyze ARs exhibiting strong field polarity reversal lines by comparing the natural clustering of these ARs to the calculated $R$ value as described in Section~\ref{sub:neutralline_method}. When we apply dimensionality reduction on this data using SVD, the resulting dictionary
is very similar to Figure~\ref{fig:pcadict} with the first two
patches consisting of uniform nonzero patches, the third and fourth
patches consisting of gradients in the magnetogram, and the fifth
and sixth patches consisting of gradients in the continuum.

As before, the mean silhouette width indicates that the appropriate
number of clusters is 2. When we cluster the ARs using the SVD coefficients
corresponding to all six patches, we obtain a correlation between
cluster assignment and $R$  of $0.09$ (see Table~\ref{tab:corr}). This isn't particularly high which suggests that the natural geometry based on the distribution of all six coefficients does not correlate well with $R$. However, since the clustering is separating the ARs based on the natural
geometry in the spaces we are considering, we can influence the
clustering by choosing the space. In other words, if we restrict our
analysis to only coefficients corresponding to specific dictionary
patches, then this will influence the clustering. 

Restricting the clustering analysis to the SVD coefficients corresponding
only to the magnetogram components (i.e. elements 2, 3, and 4 in Figure~\ref{fig:pcadict}) results in a correlation of $0.30$ between cluster assignment and $R$ value. If we only consider the gradient components (elements 3 and
4), then the correlation is $0.34$. 

The relationship between cluster assignment and $R$ may not be linear
as the correlation between the clustering assignment using only the
gradient components and $\log R$  is $0.45$. Comparing
the magnetogram only components based clustering with $\log R$ similarly
increases the correlation coefficient. Given that clustering is an
unsupervised method and that we are only clustering into two groups,
this correlation is quite high. This suggests that the natural geometry
of the image patch analysis increasingly corresponds with $R$
as we restrict the analysis to magnetogram gradients. Supervised methods,
such as regression, should lead to an even greater correspondence. Additionally, since the correlation is not perfect, this suggests that there is information in the image patch analysis that is not present in the $R$ value which may be useful for AR analysis.

For NMF, when we include all of the coefficients,
the correlation between $R$  and clustering assignment is $0.15$.
While this is small, we are again comparing the labels of an unsupervised
approach to a continuum of values. Thus we can expect that the performance
would be better in a supervised setting. If we compare the clusering to $\log R$, the correlation decreases to $0.08$. It is difficult to
restrict the NMF dictionary to only continuum and magnetogram parts
and gradients as most of the components contain a gradient component
in the magnetogram. Therefore we only cluster the ARs using all NMF coefficients.

\begin{table}

\caption{Magnitude of the correlation coefficient of the clustering assignment
with either $R$ or $\log R$ when using all of the coefficients,
only the coefficients corresponding to the magnetogram component (SVD
elements 2-4 in Figure 4), or only magnetogram gradient coefficients
(SVD elements 3-4 in Figure 4). For NMF, all of the dictionary elements
are associated with the magnetogram and many of them have gradient
components so we only perform clustering with all of the coefficients.\label{tab:corr}}
\centering

\begin{tabular}{|c|c|c|c|c|}
\cline{2-5} 
\multicolumn{1}{c|}{} & \multicolumn{3}{c|}{SVD} & \multicolumn{1}{c|}{NMF}\tabularnewline
\hline 
 & All & Mag. only  & Grad. only & All\tabularnewline
\hline 
$R$ & $0.09$ & $0.30$ & $0.34$ & $0.15$\tabularnewline
\hline 
$\log R$ & $0.02$ & $0.37$ & $0.45$ & $0.08$\tabularnewline
\hline
\end{tabular}
\end{table}

\section{Conclusion}

In this work, we introduce a reduced dimension representation  of an AR that allows a data-driven unsupervised classification of ARs based on their local geometry. The ROI that surrounds and includes the AR represents its most salient part and must be provided by the user. We used STARA masks in conjunction with  masks situated around the neutral line, and compared our results with the Mount Wilson classification in order to ease interpretation of the unsupervised scheme.  

The Mount Wilson scheme focuses on the largest length scale when describing the geometrical
arrangements of the magnetic field, whereas our method focuses on classifying
ARs using information from fine length scale.  We have shown that when we analyze and
cluster the ARs based on the global statistics of the local properties,
there are similarities to the classification based on the large scale
characteristics. For example, when clustering using the Hellinger distance, one cluster contained most of the complex ARs. Other large
scale properties such as the size of the AR also influenced the clustering
results. Table~\ref{tab:recap} summarizes the properties that are found to influence the clustering under the two schemes.

\begin{table}
\caption{Summary of features distinguishing the clusters under the various classification schemes tested.\label{tab:recap}}
\centering

\begin{tabular}{|l|l|l|}
\hline 
Class. Scheme & Cluster 1 & Cluster 2\tabularnewline
\hline
\hline 
SVD Hellinger & largest $\beta,\beta\gamma$ sunspots; majority of $\beta\gamma\delta$; & smallest $\beta,\beta\gamma$ sunspots; high \tabularnewline
& high concentration of  patches with  & concentration of   patches with average 
\tabularnewline
& average magnetic field value $\simeq 0$; & magnetic field value close to $+1$ or $-1$; \tabularnewline
 &  large Hellinger distance between ARs & small Hellinger distance between ARs \tabularnewline
\hline
NMF Hellinger &  largest $\alpha,\beta,\beta\gamma$ sunspots; majority of $\beta\gamma\delta$;& smallest $\alpha,\beta,\beta\gamma$ sunspots; \tabularnewline
 & large Hellinger distance between ARs  & small Hellinger distance between ARs \tabularnewline
\hline

\end{tabular}

\end{table}

In this comparison with the Mount Wilson scheme, we  found  that the STARA masks were sometimes too restrictive
 which led to a mismatch between the Mount Wilson label
and the extracted data. For example, there were several cases where
an AR was labeled as a $\beta$ class but the STARA mask only extracted
magnetic field values of one polarity.  We showed that the neutral line contains additional information about the complexity of the AR. For this reason, we expect that including information beyond the STARA masks will lead to improved
matching with the Mount Wilson labels. 

To investigate the possibility for our method to distinguish between potential  and non-potential fields, we  considered a ROI made of pixels situated along high-gradient, strong field polarity reversal lines. This is the same ROI as that used in the computation of the $R$ value, which has proved useful in flare prediction in a supervised context. We found that our clustering was correlated with the $R$ value, that is, the clustering based on the reduced dimension representation separates ARs corresponding to low $R$ from the ones with large $R$. 

In future work, we plan to study the efficiency of supervised techniques applied to the reduced dimension representation.
 Supervised classification 
can always do at least as well as unsupervised learning in the task of reproducing class labels (e.g. Mount Wilson label). Indeed, in supervised classification, a training dataset with labels must be provided. A classifier (or predictor) is then built within the input feature space, and is used to provide a label to new observations. In contrast, unsupervised classification or clustering separates the ARs naturally based on the geometry of the input feature space and does not use labels.
Thus if the goal is for example to reproduce the Mount Wilson classes,  or to detect nonpotentiality using global statistics of local properties, then supervised methods would lead to increased correspondence relative to our unsupervised results.

In case of flare prediction, the labels would be some indicator of flare activity such as the strength of the largest flare that occurred within a specified time period after the image was taken. Supervised techniques such as classification or regression could be applied depending on the nature of the label (i.e. categorical vs. continuum).

A good way of assessing how well a given feature space can do in a
supervised setting is to estimate the Bayes error. The Bayes error
gives the minimum average probability of error that any classifier
can achieve on the given data and can be estimated in the two class
setting by estimating upper and lower bounds such as the Chernoff
bound using a divergence-based estimator as in \cite{moon2014nips}.
These bounds can be estimated using various schemes and combinations
of data (inside sunspots, along the neutral line, etc.) to determine
which scheme is best at reproducing the desired labels (e.g., the
Mount Wilson labels). This can also be done in combination with physical parameters of the ARs such those used in \cite{bobra2015solar} (e.g. the total unsigned flux, the total area, the sum of flux near the polarity inversion line, etc.).

These methods of comparing AR images can also be adapted to a time series
of image pairs. For example, image pairs from a given point in time
may be compared to the image pairs from an earlier period to measure
how much the ARs have changed. The evolution of an AR may also be studied by defining class labels based on the results from one of the clustering schemes in this paper. From the clustering results, a classifier may be trained that is then used to assign an AR to one of these clusters at each time step. The evolution of the AR's cluster assignment can then be examined.

\begin{acknowledgements}

This work was partially supported by the US National Science Foundation
(NSF) under grant CCF-1217880 and a NSF Graduate Research Fellowship
to KM under Grant No. F031543. VD acknowledges support from the Belgian
Federal Science Policy Office through the ESA-PRODEX program, grant
No. 4000103240, while RDV acknowledges support from the BRAIN.be program
of the Belgian Federal Science Policy Office, contract No. BR/121/PI/PREDISOL.
We thank Dr. Laura Balzano at the University of Michigan, as well as Dr. Laure Lef\`{e}vre, Dr. Raphael Attie, and Ms. Marie Dominique at the Royal Observatory of Belgium for their feedback
on the manuscript. The authors gratefully acknowledge Dr. Paul Shearer's
help in implementing the modified NMF algorithm. The editor thanks two anonymous referees for their assistance in evaluating this paper.

\end{acknowledgements}

\begin{appendix}

\section{Method Details}

\subsection{Matrix Factorization}

\label{sub:dict_appendix}As mentioned in Section~\ref{sec:dictionary},
the goal of matrix factorization is to accurately decompose
the $2m^{2}\times n$ data matrix $\mathbf{Z}$ into the product of
two matrices $\mathbf{A}$ (with size $2m^{2}\times r$) and $\mathbf{H}$
(with size $r\times n$), where $\mathbf{A}$ has fewer columns than
rows ($r<2m^{2}$). The matrix \textbf{$\mathbf{A}$} is the dictionary
and the matrix \textbf{$\mathbf{H}$} is the coefficient matrix. The
columns of $\mathbf{A}$ form a basis for the data in $\mathbf{Z}$. 

The two matrix factorization methods we use are singular value decomposition (SVD) and nonnegative matrix factorization (NMF). These two
methods can be viewed as solving two different optimization problems
where the objective function is the same but the constraints differ.
Let $\mathbf{A}=[\mathbf{a}_{1},\mathbf{a}_{2},\dots,\mathbf{a}_{r}]$
and $\mathbf{H}=[\mathbf{h}_{1},\mathbf{h}_{2},\dots\mathbf{h}_{n}]$.
For SVD, the optimization problem is \[
\begin{array}{cc}
\min_{\mathbf{A},\mathbf{H}} & ||\mathbf{Z}-\mathbf{AH}||_{F}^{2}\\
\text{subject to} & \mathbf{a}_{i}^{T}\mathbf{a}_{j}=\begin{cases}
1, & i=j\\
0, & i\neq j\end{cases}\end{array}.\]
In words, SVD requires the columns of $\mathbf{A}$ to be orthonormal.

For standard NMF, the optimization problem is \[
\begin{array}{ccc}
\min_{\mathbf{A},\mathbf{H}} & ||\mathbf{Z}-\mathbf{AH}||_{F}^{2}\\
\text{subject to} & \mathbf{a}_{i}\geq0, & \forall i=1,\dots,r\\
 & \mathbf{h}_{i}\geq0, & \forall i=1,\dots,n\end{array},\]
where $\mathbf{a}\geq0$ applied to a vector $\mathbf{a}$ implies
that all of $\mathbf{a}$'s entries are greater than or equal to 0.
In our problem, only the continuum is nonnegative so we only apply
the constraint to the continuum part of the matrix $\mathbf{A}$.
So if $\mathbf{a}_{i}$ and $\mathbf{b}_{i}$ are both vectors with
length $m^{2}$ corresponding to the continuum and magnetogram parts,
respectively, then we have $\mathbf{A}=\left[\begin{array}{cccc}
\mathbf{a}_{1} & \mathbf{a}_{2} & \dots & \mathbf{a}_{r}\\
\mathbf{b}_{1} & \mathbf{b}_{2} & \dots & \mathbf{b}_{r}\end{array}\right]$. The NMF method we use also constrains the columns of $\mathbf{H}$
to lie on a simplex,\textbf{ }i.e. $\sum_{j=1}^{r}\mathbf{h}_{i}(j)=1$.
Thus the optimization problem for our approach to NMF is \[
\begin{array}{ccc}
\min_{\mathbf{A},\mathbf{H}} & ||\mathbf{Z}-\mathbf{AH}||_{F}^{2}\\
\text{subject to} & \mathbf{a}_{i}\geq0, & \forall i=1,\dots,r\\
 & \mathbf{h}_{i}\geq0, & \forall i=1,\dots,n\\
 & \sum_{j=1}^{r}\mathbf{h}_{i}(j)=1, & \forall i=1,\dots,n\end{array}.\]
This problem is not convex and is solved in an alternating manner
by fixing $\mathbf{H}$, finding the matrix $\mathbf{A}$ that solves
the problem assuming $\mathbf{H}$ is fixed, and then solving for
$\mathbf{H}$ while $\mathbf{A}$ is fixed. This process is repeated
until the algorithm converges to a local minimum. See \cite{lin2007nmf}
for more details on the convergence analysis.

\subsection{The EAC-DC Clustering Method}

\textbf{\label{sub:EACDC}}Let $V=\{v_{1},v_{2},\dots,v_{N}\}$ be
a set of vertices and let $E=\{e_{ij}\}$, where $e_{ij}$ denotes
an edge between vertices $v_{i},\, v_{j},\, i,j\in\{1,\dots,N\}$,
be a set of undirected edges between them. The pair $(V,E)=G$ is
the corresponding undirected graph. In our application, $V$ corresponds
to the set of AR image pairs being clustered and $E$ contains all
possible edges between the vertices. The weight of an edge $e_{ij}$
is defined as $w_{ij}$ and measures the base dissimilarity between
two vertices $v_{i}$ and $v_{j}$. In many applications, the base
dissimilarity is the Euclidean distance. In our case, we use the Hellinger distance as the base dissimilarity
measure.

A spanning tree $T$ of the graph $G$ is a connected acyclic subgraph
that passes through all $N$ vertices of the graph and the weight
of $T$ is the sum of all the edge weights used to construct the tree,
$\sum_{e_{ij}\in T}w_{ij}$. A minimal spanning tree of $G$ is a
spanning tree which has the minimal weight $\min_{T}\sum_{e_{ij}\in T}w_{ij}$.

Prim's algorithm \citep{prim1957shortest} is used by \cite{galluccio2013clustering}
to construct the dual rooted MST. In Prim's algorithm, the MST is
grown sequentially where at each step, a single edge is added. This
edge corresponds to the edge with minimal weight that connects a previously
unconnected vertex to the existing tree. The root of the MST corresponds
to the beginning vertex. For the dual rooted MST, we begin with two
vertices $v_{i}$ and $v_{j}$ and construct the minimal spanning
trees $T_{i}$ and $T_{j}$. At each step, the two edges that would
grow both trees $T_{i}$ and $T_{j}$ using Prim's algorithm are proposed
and the edge with minimal weight is added. This continues until $T_{i}$
and $T_{j}$ connect. The weight of the final edge added in this algorithm
defines a new metric between the vertices $v_{i}$ and $v_{j}$. This
process is repeated for all pairs of vertices and this new metric
is used as input to spectral clustering \citep{galluccio2013clustering}.

\begin{figure}
\centering

\includegraphics[width=0.34\textwidth]{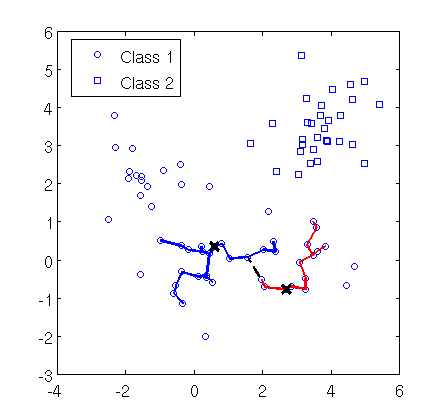}\includegraphics[width=0.34\textwidth]{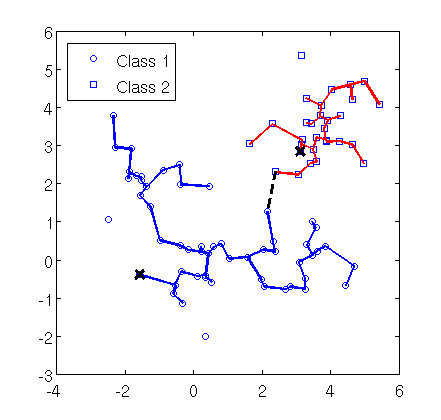}

\caption{Dual rooted Prim tree built on a 2-dimensional data set when the roots
are chosen from the same class (left) and different classes (right).
The X's mark the roots of the trees and the dashed line is the last
connected edge. The length of the last connected edge is greater when
the roots belong to clusters that are more separated. \label{fig:Dual_trees}}

\end{figure}

A primary advantage of this metric based on the hitting time of the
two MSTs is that it depends on the MST topology of the data. Thus
if two vertices belong to the same cluster, then the MST distance
between them will be small since cluster points will be close together.
This is the case even if the vertices are far away from each other
(e.g. on opposite ends of the cluster). However, if the two vertices
are in different clusters that are well separated, then the MST distance
between them will be large. See Figure~\ref{fig:Dual_trees} for
an example. Thus this method of clustering is very robust to the shape
of the clusters. \cite{galluccio2013clustering} contains many more
examples. 

The MST based metric can be computationally intensive to compute as
Prim's algorithm must be run as many times as there are pairs of vertices.
To counter this, \cite{galluccio2013clustering} proposed the EAC-DC
algorithm which uses the information from only a subset of the dual
rooted MSTs. This is done by calculating the dual rooted MSTs for
a random pair of vertices. Three clusters are defined for each run:
all vertices that are connected to one of the roots in the MSTs form
two of the clusters (one for each root) while all points that are
not connected to either of the MSTs are assigned to a third {}``rejection''
cluster. A co-association measure for two vertices is then defined
as the number of times those vertices are contained in the same non-rejection
cluster divided by the total number of runs (dual rooted MSTs). This
co-association measure forms a similarity measure to which spectral
clustering is applied.

\end{appendix}
\bibliographystyle{swsc}
\bibliography{SWSC_Moon}
\end{document}